\documentclass[oneside,11pt]{llncs}
\usepackage[a4paper,hcentering,dvips]{geometry}
\usepackage{hyperref,ifpdf,color}
\usepackage{amsmath,bbm}
\usepackage[latin1]{inputenc}
\DeclareMathOperator\lsb{lsb}

\makeatletter
\def\l@author#1#2{}%
\def\l@title#1#2{}%
\ifpdf\else\let\url\HyOrg@url\fi
\makeindex

\def\marginpar#1{}
\newcommand\TODO[1][]{%
  {\toks255{#1}\typeout{*** TODO\ifx\TODO#1\TODO\else: \the\toks255\fi}}%
  \noindent\textbf{[TODO\ifx\TODO#1\TODO\else: #1\fi]}%
  \marginpar{\hbox{\textbf{TODO\ifx\TODO#1\TODO\else: #1\fi}}\hss}%
}
\newcommand\biindex[2]{\index{#1 #2}\index{#2!#1}}
\newcommand\triindex[3]{\index{#1 #2 #3}\index{#2 #3!#1}\index{#3!#1 #2}}
\let\oldindex\index
\def\index#1{\oldindex{#1}%
  \marginpar{\hbox{#1}\hss}}
\newcommand\sfTH{\ensuremath{\mathsf{TH}}}
\newcommand\sfH{\ensuremath{\mathsf H}}
\newcommand\sfA{\ensuremath{\mathsf A}}
\newcommand\sfAr{\ensuremath{\mathsf{A_{real}}}}
\newcommand\sfAs{\ensuremath{\mathsf{A_{sim}}}}
\newcommand\port[2]{\mathsf{#1}%
  \ifx#2-^\leftrightarrow\else\ifx#2<^\triangleleft\else#2\fi\fi}
\newcommand\masterclk{\port{clk}<?}
\newcommand\buff[1]{\mathsf{\tilde{#1}}}
\newcommand\calP{\mathcal P}
\newcommand\calE{\mathcal E}
\newcommand\calH{\mathcal H}
\newcommand\calI{\mathcal I}
\newcommand\calO{\mathcal O}
\newcommand\setN{\mathbbm N}
\newcommand\setC{\mathbbm C}
\newcommand\setR{\mathbbm R}
\newcommand\bra[1]{\ensuremath{\langle #1\rvert}}
\newcommand\ket[1]{\ensuremath{\lvert #1\rangle}}
\newcommand\butter[1]{\ensuremath{\ket{#1}\bra{#1}}}
\newcommand\Rho{\ensuremath{\mathrm P}}
\newcommand\Ports{\mathit{Ports}}
\newcommand\CPorts{\mathit{CPorts}}
\newcommand\Deltabuff{\Delta_{\mathrm{buff}}}
\newcommand\MCS{{\mathsf M_{CS}}}
\newcommand\run{\mathit{run}}
\newcommand\view{\mathit{view}}
\DeclareMathOperator\tr{tr}
\newcommand\free{\mathsf{free}}
\newcommand\forb{\mathsf{forb}}
\newcommand\Conf{\mathsf{Conf}}
\newcommand\conf{\mathit{conf}}
\DeclareMathOperator\SD{StatDist}
\newcommand\SEC{\mathsf{sec}}
\newcommand\Comb{\mathsf{Comb}}
\newcommand\QStates{\mathit{QStates}}
\newcommand\CStates{\mathit{CStates}}
\newcommand\Fin{\mathit{Fin}}

\begin{document}

\title{Simulatable Security for Quantum Protocols}
\author{Dominique Unruh}
\institute{IAKS, Arbeitsgruppe Systemsicherheit,\\
           Fakultät für Informatik, Universität Karlsruhe,
           Am Fasanengarten 5,\\ 76131 Karlsruhe, Germany}
\maketitle

\pagestyle{plain}

\begin{abstract}
  The notion of simulatable security (reactive simulatability,
  universal composability) is a powerful tool for allowing the modular
  design of cryptographic protocols (composition of protocols) and
  showing the security of a given protocol embedded in a larger one.
  Recently, these methods have received much attention in the quantum
  cryptographic community
  (e.g.~\cite{Renner:2004:Universally,Ben-Or:2004:Universal}).
  
  We give a short introduction to simulatable security in general and
  proceed by sketching the many different definitional choices
  together with their advantages and disadvantages.

  Based on the reactive simulatability modelling of Backes, Pfitzmann
  and Waidner \cite{Backes:2004:Secure} we then develop a quantum security model. By following the
  BPW modelling as closely as possible, we show that composable
  quantum security definitions for quantum protocols can strongly
  profit from their classical counterparts, since most of the
  definitional choices in the modelling are independent of the
  underlying machine model.
  
  In particular, we give a proof for the simple composition theorem
  in our framework.
  \medskip
  

\end{abstract}

\vfill

\begin{center}
  \framebox{% \parbox{5in}{
      \begin{tabular}{ll}
        \bf Version history\\
        \hline
        Version 1, 18 Sep 2004\qquad\null & Initial version \\
        Version 2, 17 Nov 2004 & Added proof of combination lemma \\
        & Added comparison to the model of \cite{Ben-Or:2004:General} \\
        & Minor corrections
      \end{tabular}
    }
\end{center}

\newpage
\renewcommand\tableofcontents{%
    \section*{\contentsname
        \@mkboth{%
           \MakeUppercase\contentsname}{\MakeUppercase\contentsname}}%
    \@starttoc{toc}%
    }
\tableofcontents

\section{Introduction}
\index{introduction}
\label{sec:intro}

\subsection{Overview}
\index{overview}

In Section~\ref{sec:contrib} we state what the contribution of this work is.

In Section~\ref{sec:terminology} we give a show statement about the
use of terminology in the field of simulatable security.

In Section~\ref{sec:history} we try to give an overview of the
historical development of simulatable security.

Sections~\ref{sec:necessity} and~\ref{sec:whatis} give a short
introduction to the notion of simulatable security.

Sections~\ref{sec:order.quant}--\ref{sec:sched} gives an overview on
design decisions appearing in the modelling of simulatable security.

Section~\ref{sec:bom} gives a short comparison between our model and
that of \cite{Ben-Or:2004:General}.

Section~\ref{sec:quant} very tersely recapitulates the quantum
mechanical formalism used in this work.

Section~\ref{sec:qns} is concerned with the actual definition of
quantum machines and quantum networks.

Based on these definitions, Section~\ref{sec:secdef} gives a security
definition for quantum protocols.

Section~\ref{sec:comp} introduces the definition of composition and
shows the simple composition theorem.

Finally, in Section~\ref{sec:concl} some concluding remarks can be found.

\subsection{Our contribution}
\label{sec:contrib}

Our contribution in this work is threefold:
\begin{itemize}
\item In the introduction we give survey on design decision that have
  to be made when designing a simulatable security model. We do not
  only expose the decisions involved in our definition, but try to
  look at other models, too. The problems in the definitions of
  simulatable security models are often underestimated, we hope that
  our survey will give an impression what problems lies ahead on the
  (probably still quite long) route of finding a simple and convincing
  model of security.
\item Our second contribution is to show, that when defining quantum
  security models, many of the decisions to be made are not related to
  the quantum nature of the communication, but would already appear in
  a classical modelling.

  To emphasise that point, we develop our model in strong similarity
  to the classical modelling of \cite{Backes:2004:Secure}, our quantum
  model can therefore be seen as a quantum extension of that modelling.

  By this we hope to show that a quantum and classical security models
  should be developed hand in hand. First solve the problems appearing
  in the classical modelling, and then try and lift the classical
  model to the quantum case.
\item Third we give a concrete model of security. We hope that this
  model will show problems and possibilities in simulatable security,
  and be a step on the way towards a simple and yet general security
  definition.

  The current version seems to represent a consistent modelling of
  security, however the author has to admit that the generality in the
  modelling of scheduling had a prize: complexity. We fear that in the
  present modelling a complete formal proof of security might be quite
  unwieldy. This can be seen in the present work in
  Section~\ref{sec:combi}, where the author was unable to find a
  \emph{readable} proof for the statements there and therefore took recourse
  to a rather vague sketch. We hope that future security models will
  solve this problem without loss of generality. (See also
  Section~\ref{sec:concl}).
\end{itemize}

\subsection{A word on terminology}
\index{terminology}
\label{sec:terminology}

Some confusion exists when it comes to actually finding a name for the
concept of simulatable security. In order to prevent misunderstandings
and allow the reader to compare the present modelling with others, we
will shortly comment on the terminology used in the present work.

The most widely known term is \emph{universal
composability.}\biindex{universal}{composition} This notion was
introduced by Canetti\index{Canetti}
\cite{Canetti:2001:Security}. Since then, the notion \emph{universal
composition} and especially \emph{UC
framework}\biindex{UC}{framework} became strongly associated with the
model of Canetti. However, the word \emph{universal composability} is
used in two other ways: first, it is used to denote the property to be
secure according to definitions similar to that of
\cite{Canetti:2001:Security} or \cite{Backes:2004:Secure}, without
meaning the model of Canetti in particular. Second, \emph{universal
composability} often denotes the applicability of the Composition
Theorem. However, since there are different flavours of the
Composition Theorem, sometimes \emph{universal composability} means
concurrent and simple composability\footnote{See
Section~\ref{sec:comp.flavours} for an analysis of the difference
between \emph{concurrent} and \emph{simple composability}.}  (e.g. in
\cite{Canetti:2001:Security}), while on other occasions it is used for
simple composability.

Besides the model of Canetti, another model found much interest in the
last few years: the model of Backes, Pfitzmann, and Waidner
\cite{Pfitzmann:2001:Model,Backes:2004:Secure}. While the term of
\emph{universal composability} in its most general meaning can be
applied to that model, too, Backes\index{Backes},
Pfitzmann\index{Pfitzmann}, and Waidner\index{Waidner}
\cite{Backes:2004:Secure} prefer the use of \emph{reactive
  simulatability.}\biindex{reactive}{simulatability}

In order to avoid such confusion, we will adhere to the following
convention in the present exposition: for the modelling by Canetti we
will use the term \emph{UC framework}. The modelling by BPW we shall
name \emph{reactive simulatability} (or shorter \emph{RS
  framework}\biindex{RS}{framework}), while the overall concept of
security notions using simulation and guaranteeing composition
(encompassing these two modellings) we will call \emph{simulatable
  security.}\biindex{simulatable}{security}

The different flavours of the composition theorem we will
differentiate by using the attributes \emph{simple},
\emph{concurrent}, and the combination of both (see
Section~\ref{sec:comp.flavours}). In this nomenclature the \emph{Universal
Composition Theorem} of \cite{Canetti:2001:Security} would be called
\emph{Simple and Concurrent Composition Theorem}, while the
\emph{Composition Theorem} of \cite{Backes:2004:Secure} would be named
\emph{Simple Composition Theorem}. We restrain from using the shorter
term \emph{universal composition} for \emph{simple and concurrent
composition} to prevent confusion with the other meanings described
above.

A further term which is noteworthy in this context is that of the
\emph{honest user} and the \emph{environment}, resp. Both denote the
same idea, the first being used in the RS framework, the second in the
UC framework.  We will use these two notions in an interchangeable
way, preferring \emph{environment} when trying to motivate or explain
on an intuitive level, while sticking to \emph{honest user} when
giving formal definitions or proofs. The same holds for the terms
\emph{trusted host} (RS framework) and \emph{ideal functionality} (UC
framework).

More terms will appear in the course of this exposition which have
different translations in different frameworks. We will mention these
when introducing the notions in our exposition. The reader is strongly
encouraged to use the index (p.~\pageref{index}) to find these
translations.

\subsection{A brief historical account}
\index{history}
\label{sec:history}

To the best of our knowledge, the notion of a simulator to define
security of a protocol was first introduced in the definition of
zero-knowledge\index{zero-knowledge} proofs
\cite{Goldwasser:1985:Knowledge}. Here the simulation paradigm was
used to ensure that the verifier could not learn anything except the
truth of the statement to be proven. This was done by requiring that
any transcript of the interaction between prover and verifier could
also be generated by the simulator (without knowing a proof witness
for the statement). If this is possible, then the interaction of
course does not allow to learn anything about the proof.

However, it turned out that this definition, while capturing very well
the idea that the verifier learns nothing about the proof, does not
guarantee the possibility to compose two zero-knowledge protocols in
parallel (without losing the zero-knowledge property)
\cite{Goldreich:1996:Composition}.

Note further, that here the simulation paradigm was only used for one
of the required properties (being zero-knowledge), the soundness
condition (i.e., that the protocol indeed is a proof) was defined in
another manner. In this it differs from today's simulatable security
which aspires to capture all security properties in one single
definition.

In \cite{Beaver:1991:Foundations} the notion of \emph{relative
  resilience}\biindex{relative}{resilience} was introduced. This
notion allowed to say that one protocol $\pi$ was at least as secure
as another protocol/trusted host $\rho$ by requesting that for all
protocol inputs and all adversaries, there is a simulator, so that the
outputs of $\pi$ with the adversary and $\rho$ running with the
simulator are indistinguishable. This was a major step towards today's
notion of simulatability, since a protocol's security was now defined
in comparison to an ideal specification.
\cite{Beaver:1991:Foundations} showed, that the security definition
was closed under sequential
composition\biindex{sequential}{composition} (executing one protocol
\emph{after} another). However, when executing protocols in parallel,
no guarantee was given.
%\footnote{In fact, one can easily construct a
%  counter-example: In the model of \cite{Beaver:1991:Foundations}
%  coin-toss as proposed in \cite{Blum:1981:Coin} was secure. However,
%  it follows from the impossibility results in
%  \cite{Canetti:2001:Commitments}, that this coin-toss can not be
%  composable.} 
At the same time \cite{Micali:1991:Secure} announced another model
based a notion of simulatability, even achieving some kind of
composition (\emph{reducibility}). It is unknown to the author,
whether this idea was further pursued.

Later \cite{Pfitzmann:2000:Secure} and \cite{Canetti:2000:Security}
introduced independent models with synchronous scheduling (see
Section~\ref{sec:sched}). The model of \cite{Pfitzmann:2000:Secure} (a
predecessor of the BPW-model \cite{Backes:2004:Secure} underlying our
model) already had a \emph{simple composition theorem}, i.e.~one could
use one protocol as a sub-protocol of another, and the sub-protocol
could run simultaneously with the calling protocol (see
Section~\ref{sec:comp.flavours}).  \cite{Canetti:2000:Security}
achieved a similar result, however the security of a composed protocol
could only be ensured in \cite{Canetti:2000:Security} if the calling
protocol was suspended until the sub-protocol terminated. When the
calling and the called protocol where executed simultaneously, no
guarantee would be made. So this composition theorem may be classified
as being a sequential composition theorem, similar (but somewhat more
powerful) to that of \cite{Beaver:1991:Foundations}.

Shortly afterwards, both Pfitzmann, Waidner
\cite{Pfitzmann:2001:Model}, and Canetti \cite{Canetti:2001:Security}
presented asynchronous versions of their models.
\cite{Pfitzmann:2001:Model} had---in our nomenclature---a \emph{simple
  composition theorem,} and \cite{Canetti:2001:Security} a
\emph{simple and concurrent composition theorem} (see
Section~\ref{sec:comp.flavours}). In
\cite{Backes:2004:GeneralComposition} it was shown, that not only simple but
also concurrent composition is also possible in their model. 

To the best of our knowledge, the first simulation based quantum
security models\biindex{quantum}{security model} were
\cite{Graaf:1998:Towards} and \cite{Smith:2001:Quantum}. Like
\cite{Beaver:1991:Foundations}, both did not have the notion of an
honest user/environment, therefore they too could only guarantee
sequential composition. The first models having an honest
user/environment were the independent works \cite{Ben-Or:2002:Quantum}
and \cite{Unruh:2002:Formal}. Both based on Canetti's model, they both
provided simple and concurrent composition (see
Section~\ref{sec:comp.flavours}).

\subsection{On the necessity of simulatable security}
\label{sec:necessity}\biindex{necessity of}{simulatable security}

A question that may arise is, why do we need another security
definition. Are there not sufficiently many definitions like
\emph{privacy}\index{privacy}, \emph{correctness}\index{correctness},
\emph{robustness}\index{robustness},
\emph{non-maleability}\index{non-maleability}, etc. (the list of
security properties\biindex{security}{properties} found in the
literature is very long)? Why add another one?

There are several good reasons for this:
\begin{itemize}
\item There are (admittedly constructed) protocols, where the privacy
  requirement is fulfilled, where the correctness requirement is
  fulfilled, but where in information nevertheless leaks. The example
  goes back to Mical and Rogaway and we cite here the version found in
  the introduction of [Graaf:1998:Towards]:
  
  Let $x$ and $y$ be the inputs of Alice and Bob, resp.
  They want to evaluate the function $g$ given by
  $$
  g(x,y) := \begin{cases}
    0x, & \text{if }\lsb(y)=0, \\
    1y, & \text{if }\lsb(y)=1,
  \end{cases}
  $$
  Here $\lsb(y)$ stands for the last bit of $y$, and $0x$ and $1y$
  for concatenation.

  Consider the following protocol for this task:
  \begin{enumerate}
  \item Alice and Bob commit to the bits of their inputs using some secure bit
    commitment scheme.
  \item Bob unveils $\lsb(y)$.
  \item If $\lsb(y)=0$, Bob unveils $x$. If $\lsb(y)=1$, Alice unveils
    $y$. Now the both parties can calculate $g(x,y)$.
  \item Additionally (just for the sake of the counterexample), Alice
    sends $x$ to Bob.
  \end{enumerate}
  
  Clearly, the protocol is correct (i.e.~the function $g(x,y)$ is
  always evaluated correctly). Further, the privacy condition is
  fulfilled, since in the specification of the protocol there always
  is a way for Bob to learn Alice's input, so the fact that $x$ is
  sent to Alice does formally not violate the privacy condition
  (see~[Graaf:1998:Towards] for details).
  
  But intuitively, the function is not evaluated in a secure fashion,
  since in the specification Bob would learn Alice's input only for
  $\lsb(y)=0$, while in the protocol he always learns it.
  
  So a new definition is needed to capture both privacy and
  correctness in one go. Simulatable security has this advantage.
\item The second problem is that the list of desirable security
  properties is ever growing, it is not restricted to just privacy and
  correctness. The simulatable security encompasses many definitions
  at once using a very general approach (see
  Section~\ref{sec:whatis}), so one can have more confidence that the
  intuitively security is guaranteed by simulatable security.  (Note
  however, that some special security properties like incoercibility are not
  guaranteed by simulatable security
  \cite{Mueller-Quade:2003:PersonalSg}.)
\item The probably most striking problem of many security properties
  is incomposability. If a protocol $\pi$ is given for some
  cryptographic primitive $X$ (say a bit commitment), and another
  protocol $\rho$ using $X$ securely accomplishes something great,
  then it would seem natural to combine $\pi$ and $\rho$ to get a
  protocol accomplishing the great thing without recurse to any
  primitives.  However, it turns out that there is no guarantee that
  the composed protocol is still secure. 
  
  To remove this lack of certainty, and to allow the modular
  construction of protocols, we need a security definition that is
  composable, i.e.~that allows design protocols for small task, and
  using these as building blocks for bigger protocols, without having
  to prove the security of the bigger protocols from scratch.
  
  Fortunately, simulatable security provides such composability (see
  Section~\ref{sec:comp.flavours}).
\end{itemize}

\subsection{What is simulatable security?}
\label{sec:whatis}
\biindex{simulatable}{security}

The basic idea is a follows: Assume, that we have some given
cryptographic application, and we can specify some reference protocol
(the \emph{ideal protocol}\biindex{ideal}{protocol}) or
\emph{trusted party}\biindex{trusted}{party} $\sfTH$, which does
implement the wanted behaviour in a secure way. Of course, we do now
have to design that trusted host, but this is usually easier than
designing the protocol, since we do not have to bother with details
like insecure communication, or who should carry out computations
etc., since $\sfTH$ can by definition be regarded as trusted.

If we then accept, that the trusted party is secure (though not
necessarily feasible), we can define that some protocol $\pi$ (the
\emph{real protocol}\biindex{real}{protocol}) is as secure as $\sfTH$,
if replacing $\sfTH$ in \emph{any} situation does not result in
\emph{any} disadvantage for \emph{any} person (except the adversary,
of course). But how do we formalise situations and how do we formalise
disadvantages? We simply introduce the concept of the
\emph{environment}\index{environment} and the \emph{(real)
  adversary}\biindex{real}{adversary}. The environment is some entity,
which does interact with the protocol and the adversary as a black box
and at the end may decide, whether something harmful has happened.
When we quantify over all possible environments (possibly restricting
the computational power), and for no environment (i.e.~in no
situation) some harm\index{harm} happened using $\pi$ which could not have
happened using $\sfTH$, too, then replacing $\sfTH$ by $\pi$ is
clearly a sensible course of action, at least it will surely do no
harm (none that could be detected by the environment, at least), thus
$\pi$ can be considered to be secure.

The preceding explanation still has some great drawback: An adversary
(for which the protocol's internals are no black box, of course) could
simply vary its output depending on whether we use $\pi$ or $\sfTH$.
Then the environment might simply consider the output ``$\pi$ is in
use'' as harmful, and suddenly $\pi$ would be insecure. But, when
considering our requirement, that \emph{everything harmful, which can
  happen using $\pi$, can also happen using $\sfTH$,} we can
reasonably interpret it as \emph{everything harmful, which can happen
  using $\pi$ with some adversary $\sfAr$, can also happen using
  $\sfTH$ with some (other) adversary $\sfAs$, the
  \emph{simulator}\index{simulator}.} If we accept this formulation
that implies, that in any situation the adversary may freely choose
its strategy to be as ``harmful'' as possible, we get the security
definition which we will develop and examine in this work.

One simplification we may still introduce: So far we have required,
that using $\pi$, we get \emph{at most} as much harm as when using
$\sfTH$. In fact, we can change this definition so that we require
that we get \emph{the same amount} of harm. This simplifies the
definition and is in fact equivalent, since if one environment defines
something as harm, then some other environment could define the
opposite as harm, such bounding the amount of harm from both sides and
resulting in a claim of equality. After introducing this
simplification, the word ``harm'' is of course inadequate, so we will
use the more neutral notion of the \emph{output of the environment.}
Further it is a matter of taste, whether we should say that the
environment outputs exactly one bit, one may also allow the
environment to generate a whole stream of output, which should then be
indistinguishable in runs with the real protocol and the trusted host (ideal
protocol), see Section~\ref{sec:view.bits}.

So if we put these considerations together, we get a first definition
sketch, that runs in the following lines: For any adversary $\sfAr$,
there is some adversary $\sfAs$, such that for any environment $\sfH$,
the output of $\sfH$ when running with $\sfAr$ and $\pi$ is
indistinguishable from that of $\sfH$ running with $\sfAs$ and
$\sfTH$.

In the RS framework of \cite{Pfitzmann:2001:Model,Backes:2004:Secure},
the environment is instead called the \emph{honest user}, reflecting
the intuition that its view represents anything (e.g.~any harm) that
an honest user of the protocol may experience. Further instead of the
notion \emph{trusted host}, it is also very common to speak of an
\emph{ideal functionality}\biindex{ideal}{functionality}, e.g.~in the
model of \cite{Canetti:2001:Security}.

Of course, the above definition sketch still leaves many open
questions, like what a protocol formally is, what indistinguishable
means, how messages and machines are scheduled, etc. We will try and
discuss these questions in the following paragraphs.

\subsection{On the order of quantifiers}
\label{sec:order.quant}
\biindex{order of}{quantifiers}

In Section~\ref{sec:whatis} we ``defined'' simulatable security
approximately as follows: For all adversaries there is a simulator
s.t.~for all environments the real and ideal protocol are
indistinguishable, in symbols:
$\forall\sfAr\exists\sfAs\forall\sfH\dots$. However, rereading our
motivation one might ask whether the following order of quantifiers
does not have as much justification:
$\forall\sfAr\forall\sfH\exists\sfAs\dots$, i.e.~should the simulator
be allowed to depend on the environment/honest user or not?

In fact, both order are common. \cite{Backes:2004:Secure} calls
security with respect to the
$\forall\sfAr\exists\sfAs\forall\sfH$-ordering \emph{universal
  security,}\biindex{universal}{security} and security with respect to
the $\forall\sfAr\forall\sfH\exists\sfAs$-ordering \emph{standard
  security.}\biindex{standard}{security} The model of
\cite{Canetti:2001:Security} uses the stricter notion of universal
security, while the \cite{Backes:2004:Secure} model defines both
notions.

One might wonder, whether standard and universal security are
equivalent. It was however shown in \cite{Hofheinz:2004:Comparing},
that for statistical and polynomial security (see
Section~\ref{sec:view.bits}) there are examples separating these two
notions.

There is also a very practical point separating these two notions:
while for showing a simple composition theorem (see
Section~\ref{sec:comp.flavours}), it is sufficient to have standard
security. However all proof for concurrent composition theorems known
to the author need universal security. Note that to the best of our
knowledge it is unknown whether concurrent composability could be
proven using standard simulatability.

A further ordering of quantifiers appears in
\cite{Backes:2004:Secure}: \emph{black-box
  security}\biindex{black-box}{security}. This notion, which is even
stricter than universal security, is roughly defined as
follows: There is a simulator s.t.~for all real adversaries and all
environments, the real protocol running with environment and honest
user is indistinguishable from the ideal protocol running with
environment and simulator, where the simulator has access to the
adversary in a black-box manner. In other words, the simulator is not
allowed to depend on the real adversary in any manner, but only by
using the adversary as a black-box (without rewinding). Since no
properties are known to the author which give an advantage of
black-box over universal security, we will ignore this notion in the
present work, concentrating on universal and standard security.

\subsection{Views versus output bits - the verdict of the honest user}
\label{sec:view.bits}
\index{view}\biindex{output}{bit}

In Section~\ref{sec:whatis} we required that the views or outputs of the
environment is the indistinguishable in the run of the real and the
ideal protocol. We will now discuss this indistinguishability in more
detail.

The first definitional choice to be made is whether the environment
outputs only one bit (as done in \cite{Canetti:2001:Security}) or has
a continuous stream of output, the view of the environment (as done in
\cite{Backes:2004:Secure}).  We follow the choice of
\cite{Backes:2004:Secure} and define the view of the honest user to be
the transcript of all its classical in- and outputs.

We can now define probability distributions $\mathrm{Real}_k$,
$\mathrm{Ideal}_k$ for the view/output bit in the run of the real or
ideal protocol, indexed by the security parameter.

We sketch the following major notions of indistinguishability:
\begin{itemize}
\item \emph{Perfect.}\biindex{perfect}{security}
  The distributions $\mathrm{Real}_k$ and
  $\mathrm{Ideal}_k$ are identical. If further environment, adversary
  and simulator are computationally unbounded, we talk of perfect
  security.
\item \emph{Statistical.}\biindex{statistical}{security} There is a negligible
  function\footnote{There are different possibilities what functions
    are accepted as negligible ones. Usually a function is called
    negligible if it asymptotically gets smaller than $1/p$ for any
    polynomial $p$.} $\nu$, s.t.~the statistical
  distance\footnote{Intuitively the statistical distance describes how
    good an optimal test can distinguish between two distributions.}
  of $\mathrm{Real}_k$ and $\mathrm{Ideal}_k$ is bounded by $\nu(k)$
  for all security parameters $k$.
  
  If further environment, adversary and simulator are computationally
  unbounded, we talk of \emph{strict statistical security.}\triindex{strict}{statistical}{security}
  
  We will only use the notion of strict statistical security in the
  present work.
  
  Another variant of security using the statistical
  indistinguishability is that of \emph{statistical security} as
  defined in \cite{Backes:2004:Secure}. Here we require that for any
  polynomial $l$ the prefixes of length $l(k)$ of $\mathrm{Real}_k$
  and $\mathrm{Ideal}_k$ are indistinguishable. However the simple
  composition theorem does not hold using this security notion
  \cite{Hofheinz:2004:Artefacts}.
\item \emph{Computational.}\biindex{computational}{security}
  For any algorithm $D$ (the distinguisher)
  that is polynomial-time in $k$ and has one-bit output, the outputs
  of $D$ given $\mathrm{Real}_k$ resp. $\mathrm{Ideal}_k$ as input is
  statistically indistinguishable. When restricting honest user,
  adversary and simulator to be computationally bounded (see
  Section~\ref{sec:comp.sec}) we talk of computational
  indistinguishability. A more detailed definition can be found
  e.g.~in \cite{Backes:2004:Secure}.
  
  In the case of one-bit output of the environment, computational and
  statistical indistinguishability obviously coincide.
  
  For quantum protocols, computational indistinguishability have to be
  redefined to incorporate the power of quantum distinguishers.
\end{itemize}

\subsection{Flavours of composition}
\label{sec:comp.flavours}
\index{composition}

As already stated, one of the major advantages of simulatable security
is the possibility of composition.  A composition theorem states
roughly the following: suppose we are given a protocol $\pi$
implementing a trusted host $\sfTH$, and we have another protocol
$\rho$ using $\sfTH$ as a primitive that implements some other trusted
host $\mathsf{CPLX}$. Then protocol $\rho$ using $\pi$ also implements
$\mathsf{CPLX}$. As a formula (where $\geq$ means \emph{implements}):
\begin{equation}\label{eq:comp.draft}
  \pi \geq \sfTH
  \quad\text{and}\quad
  (\rho\text{ using }\sfTH) \geq \mathsf{CPLX}
  \quad\Longrightarrow\quad
  (\rho\text{ using }\pi) \geq \mathsf{CPLX}
\end{equation}

However, it turns out that this formula can be interpreted in two
different ways. To illustrate this, we give two examples of composition.
\begin{itemize}
\item Assume that $\pi$ is a protocol implementing a key exchange
  (e.g.~\cite{Bennett:1984:Quantum}), i.e.~$\pi\geq\mathsf{KE}$.
  Assume further that $\rho$ that generates \emph{one} key using
  $\mathsf{KE}$ and then uses this key to implement a secure message
  transmission $\mathsf{SMT}$ (applying some key-reuse-strategies).
\item Assume that $\pi$ again implements key exchange. Assume further
  that $\tilde\rho$ is a protocol that implements a secure channel
  approximately as follows: For each message to be sent, a new key
  exchange is invoked. The resulting key is then used for transmitting
  an encrypted and authenticated message (e.g. using a one-time-pad
  and authentication \cite{Raub:2004:Security}).
\end{itemize}

We are now tempted to say, that in both examples we can
use~\eqref{eq:comp.draft} to show
$$
(\rho\text{ using }\pi)\geq \mathsf{SMT}
\qquad\text{and}\qquad
(\tilde\rho\text{ using }\pi)\geq \mathsf{SMT}.
$$
But considering the examples in more detail, we note that there is
a very important difference: While $\rho$ uses \emph{only one} copy of
$\mathsf{KE}$, the protocol $\tilde\rho$ employs \emph{many} copies of
$\mathsf{KE}$ (when we assume the environment to be limited to a
polynomial amount of messages (in the security parameter $k$), we can
say that at most polynomially many invocations of $\mathsf{KE}$ are
performed). So whether we can apply \eqref{eq:comp.draft} in the
second example depends on whether ``$\rho$ using $\sfTH$'' means
``using at most \emph{one} copy of $\sfTH$'' or ``using at most
\emph{polynomially many} copies of $\sfTH$.''

To be able to differentiate more clearly between these two notions of
``using'', we introduce the following conventions: Let $\rho^X$ denote
$\rho$ using \emph{one} copy of $X$. Let further denote $\sfTH^*$ the
machine simulating arbitrarily many copies of $\sfTH$, and let
$\rho^*$ denote the machine simulating arbitrarily many copies of
$\rho$. Then we can
state the following two variants of the composition theorem:
\begin{alignat}{2}
  &\text{\emph{simple:}}\quad
  && \pi \geq \sfTH
  \quad\text{and}\quad
  \rho^\sfTH \geq \mathsf{CPLX}
  \quad\Longrightarrow\quad
  \rho^\pi \geq \mathsf{CPLX}
  \label{eq:comp.simp}
\intertext{and}
&\text{\emph{concurrent:}}\quad
&& \pi \geq \sfTH
  \quad\Longrightarrow\quad
  \pi^* \geq \sfTH^*
  \label{eq:comp.concur}
\end{alignat}

Clearly, for the first example it is sufficient to use the
\emph{simple composition theorem}\biindex{simple}{composition}
\eqref{eq:comp.simp}. To prove security of the protocol constructed in the
second example, we would proceed as follows: first, by the
\emph{concurrent composition theorem}\biindex{concurrent}{composition}
\eqref{eq:comp.concur} we get $\pi^*\geq\mathsf{KE}^*$.  Then, our
assumption that $\tilde\rho$ using $\mathsf{KE}$ implements
$\mathsf{SMT}$ must be written more concretely as
$\tilde\rho^{\mathsf{KE}^*}\geq\mathsf{SMT}$. Now we can apply the
simple composition theorem \eqref{eq:comp.simp} (substituting $\pi$ by
$\pi^*$ and $\sfTH$ by $\mathsf{KE}^*$) and get
$\tilde\rho^{\pi^*}\geq\mathsf{SMT}$.

So we have seen that for composition as in the first example, simple
composition is sufficient, while for the second example simple and
concurrent composition is needed.

The ``Universal Composition Theorem'' from
\cite{Canetti:2001:Security} provides simple and concurrent
composition, the ``Secure Two-system Composition Theorem'' from
\cite{Backes:2004:Secure} only provides simple composition, but is
supplemented by the ``General Composition Theorem'' from
\cite{Backes:2004:GeneralComposition}.

It turns out that a simple composition theorem can be shown for
standard and for universal security (cf.~Section~\ref{sec:order.quant}
for an introduction of these notions). To the best of our knowledge,
no proof of the concurrent composition theorem that does not need
universal security has been published so far. Here we will  only
present the proof sketch of the simple composition theorem, for a
sketch of the concurrent composition theorem the reader may refer to
\cite{Canetti:2001:Security}.

\subsubsection{Proof sketch for the simple composition theorem.}
\triindex{proof sketch for}{simple}{composition theorem}

\begin{figure}[thb]
  \begin{center}
    \setlength{\unitlength}{4144sp}%
\begingroup\makeatletter\ifx\SetFigFont\undefined%
\gdef\SetFigFont#1#2#3#4#5{%
  \reset@font\fontsize{#1}{#2pt}%
  \fontfamily{#3}\fontseries{#4}\fontshape{#5}%
  \selectfont}%
\fi\endgroup%
\begin{picture}(3444,4148)(439,-3927)
\thinlines
{\color[rgb]{0,0,0}\put(451,-519){\framebox(458,458){}}
}%
{\color[rgb]{0,0,0}\put(1253,-519){\framebox(458,458){}}
}%
{\color[rgb]{0,0,0}\put(451,-1321){\framebox(458,458){}}
}%
{\color[rgb]{0,0,0}\put(1253,-1321){\framebox(458,458){}}
}%
{\color[rgb]{0,0,0}\put(909,-1092){\line( 1, 0){344}}
}%
{\color[rgb]{0,0,0}\put(680,-519){\line( 0,-1){344}}
}%
{\color[rgb]{0,0,0}\put(909,-267){\line( 1, 0){344}}
}%
{\color[rgb]{0,0,0}\put(1482,-519){\line( 0,-1){342}}
}%
{\color[rgb]{0,0,0}\put(909,-519){\line( 1,-1){344}}
}%
{\color[rgb]{0,0,0}\put(909,-863){\line( 1, 1){344}}
}%
\put(604,-343){\makebox(0,0)[lb]{\smash{\SetFigFont{12}{14.4}{\rmdefault}{\mddefault}{\updefault}{\color[rgb]{0,0,0}$\sfH$}%
}}}
\put(1414,-343){\makebox(0,0)[lb]{\smash{\SetFigFont{12}{14.4}{\rmdefault}{\mddefault}{\updefault}{\color[rgb]{0,0,0}$\rho$}%
}}}
\put(1414,-1152){\makebox(0,0)[lb]{\smash{\SetFigFont{12}{14.4}{\rmdefault}{\mddefault}{\updefault}{\color[rgb]{0,0,0}$\pi$}%
}}}
\put(519,-1140){\makebox(0,0)[lb]{\smash{\SetFigFont{12}{14.4}{\rmdefault}{\mddefault}{\updefault}{\color[rgb]{0,0,0}$\sfAr$}%
}}}
{\color[rgb]{0,0,0}\put(2476,-519){\framebox(458,458){}}
}%
{\color[rgb]{0,0,0}\put(3278,-519){\framebox(458,458){}}
}%
{\color[rgb]{0,0,0}\put(2476,-1321){\framebox(458,458){}}
}%
{\color[rgb]{0,0,0}\put(3278,-1321){\framebox(458,458){}}
}%
{\color[rgb]{0,0,0}\put(2934,-1092){\line( 1, 0){344}}
}%
{\color[rgb]{0,0,0}\put(2705,-519){\line( 0,-1){344}}
}%
{\color[rgb]{0,0,0}\put(2934,-267){\line( 1, 0){344}}
}%
{\color[rgb]{0,0,0}\put(3507,-519){\line( 0,-1){342}}
}%
{\color[rgb]{0,0,0}\put(2934,-519){\line( 1,-1){344}}
}%
{\color[rgb]{0,0,0}\put(2934,-863){\line( 1, 1){344}}
}%
\put(2629,-343){\makebox(0,0)[lb]{\smash{\SetFigFont{12}{14.4}{\rmdefault}{\mddefault}{\updefault}{\color[rgb]{0,0,0}$\sfH$}%
}}}
\put(3439,-343){\makebox(0,0)[lb]{\smash{\SetFigFont{12}{14.4}{\rmdefault}{\mddefault}{\updefault}{\color[rgb]{0,0,0}$\rho$}%
}}}
\put(3439,-1152){\makebox(0,0)[lb]{\smash{\SetFigFont{12}{14.4}{\rmdefault}{\mddefault}{\updefault}{\color[rgb]{0,0,0}$\pi$}%
}}}
\put(2544,-1140){\makebox(0,0)[lb]{\smash{\SetFigFont{12}{14.4}{\rmdefault}{\mddefault}{\updefault}{\color[rgb]{0,0,0}$\sfAr$}%
}}}
{\color[rgb]{0,0,0}\put(2341,-601){\dashbox{57}(1530,810){}}
}%
\put(2431, 29){\makebox(0,0)[lb]{\smash{\SetFigFont{12}{14.4}{\rmdefault}{\mddefault}{\updefault}{\color[rgb]{0,0,0}$\sfH_\rho$}%
}}}
\put(983,-3878){\makebox(0,0)[lb]{\smash{\SetFigFont{12}{14.4}{\rmdefault}{\mddefault}{\updefault}{\color[rgb]{0,0,0}(d)}%
}}}
\put(991,-1674){\makebox(0,0)[lb]{\smash{\SetFigFont{12}{14.4}{\rmdefault}{\mddefault}{\updefault}{\color[rgb]{0,0,0}(a)}%
}}}
{\color[rgb]{0,0,0}\put(2341,-2851){\dashbox{57}(1530,810){}}
}%
{\color[rgb]{0,0,0}\put(451,-2769){\framebox(458,458){}}
}%
{\color[rgb]{0,0,0}\put(1253,-2769){\framebox(458,458){}}
}%
{\color[rgb]{0,0,0}\put(451,-3571){\framebox(458,458){}}
}%
{\color[rgb]{0,0,0}\put(1253,-3571){\framebox(458,458){}}
}%
{\color[rgb]{0,0,0}\put(909,-3342){\line( 1, 0){344}}
}%
{\color[rgb]{0,0,0}\put(680,-2769){\line( 0,-1){344}}
}%
{\color[rgb]{0,0,0}\put(909,-2517){\line( 1, 0){344}}
}%
{\color[rgb]{0,0,0}\put(1482,-2769){\line( 0,-1){342}}
}%
{\color[rgb]{0,0,0}\put(909,-2769){\line( 1,-1){344}}
}%
{\color[rgb]{0,0,0}\put(909,-3113){\line( 1, 1){344}}
}%
{\color[rgb]{0,0,0}\put(2476,-2769){\framebox(458,458){}}
}%
{\color[rgb]{0,0,0}\put(3278,-2769){\framebox(458,458){}}
}%
{\color[rgb]{0,0,0}\put(2476,-3571){\framebox(458,458){}}
}%
{\color[rgb]{0,0,0}\put(3278,-3571){\framebox(458,458){}}
}%
{\color[rgb]{0,0,0}\put(2934,-3342){\line( 1, 0){344}}
}%
{\color[rgb]{0,0,0}\put(2705,-2769){\line( 0,-1){344}}
}%
{\color[rgb]{0,0,0}\put(2934,-2517){\line( 1, 0){344}}
}%
{\color[rgb]{0,0,0}\put(3507,-2769){\line( 0,-1){342}}
}%
{\color[rgb]{0,0,0}\put(2934,-2769){\line( 1,-1){344}}
}%
{\color[rgb]{0,0,0}\put(2934,-3113){\line( 1, 1){344}}
}%
\put(2431,-2221){\makebox(0,0)[lb]{\smash{\SetFigFont{12}{14.4}{\rmdefault}{\mddefault}{\updefault}{\color[rgb]{0,0,0}$\sfH_\rho$}%
}}}
\put(2011,-743){\makebox(0,0)[lb]{\smash{\SetFigFont{12}{14.4}{\rmdefault}{\mddefault}{\updefault}{\color[rgb]{0,0,0}=}%
}}}
\put(2004,-2986){\makebox(0,0)[lb]{\smash{\SetFigFont{12}{14.4}{\rmdefault}{\mddefault}{\updefault}{\color[rgb]{0,0,0}=}%
}}}
\put(604,-2593){\makebox(0,0)[lb]{\smash{\SetFigFont{12}{14.4}{\rmdefault}{\mddefault}{\updefault}{\color[rgb]{0,0,0}$\sfH$}%
}}}
\put(1414,-2593){\makebox(0,0)[lb]{\smash{\SetFigFont{12}{14.4}{\rmdefault}{\mddefault}{\updefault}{\color[rgb]{0,0,0}$\rho$}%
}}}
\put(2629,-2593){\makebox(0,0)[lb]{\smash{\SetFigFont{12}{14.4}{\rmdefault}{\mddefault}{\updefault}{\color[rgb]{0,0,0}$\sfH$}%
}}}
\put(3439,-2593){\makebox(0,0)[lb]{\smash{\SetFigFont{12}{14.4}{\rmdefault}{\mddefault}{\updefault}{\color[rgb]{0,0,0}$\rho$}%
}}}
\put(3349,-3402){\makebox(0,0)[lb]{\smash{\SetFigFont{12}{14.4}{\rmdefault}{\mddefault}{\updefault}{\color[rgb]{0,0,0}$\sfTH$}%
}}}
\put(1332,-3402){\makebox(0,0)[lb]{\smash{\SetFigFont{12}{14.4}{\rmdefault}{\mddefault}{\updefault}{\color[rgb]{0,0,0}$\sfTH$}%
}}}
\put(527,-3390){\makebox(0,0)[lb]{\smash{\SetFigFont{12}{14.4}{\rmdefault}{\mddefault}{\updefault}{\color[rgb]{0,0,0}$\sfAs$}%
}}}
\put(2559,-3390){\makebox(0,0)[lb]{\smash{\SetFigFont{12}{14.4}{\rmdefault}{\mddefault}{\updefault}{\color[rgb]{0,0,0}$\sfAs$}%
}}}
\put(3601,-1682){\makebox(0,0)[lb]{\smash{\SetFigFont{12}{14.4}{\rmdefault}{\mddefault}{\updefault}{\color[rgb]{0,0,0}$\approx$}%
}}}
\put(3026,-3878){\makebox(0,0)[lb]{\smash{\SetFigFont{12}{14.4}{\rmdefault}{\mddefault}{\updefault}{\color[rgb]{0,0,0}(c)}%
}}}
\put(3034,-1674){\makebox(0,0)[lb]{\smash{\SetFigFont{12}{14.4}{\rmdefault}{\mddefault}{\updefault}{\color[rgb]{0,0,0}(b)}%
}}}
\end{picture}
  \end{center}
  \caption{Simple composition theorem}
  \label{fig:proof.sc}
\end{figure}
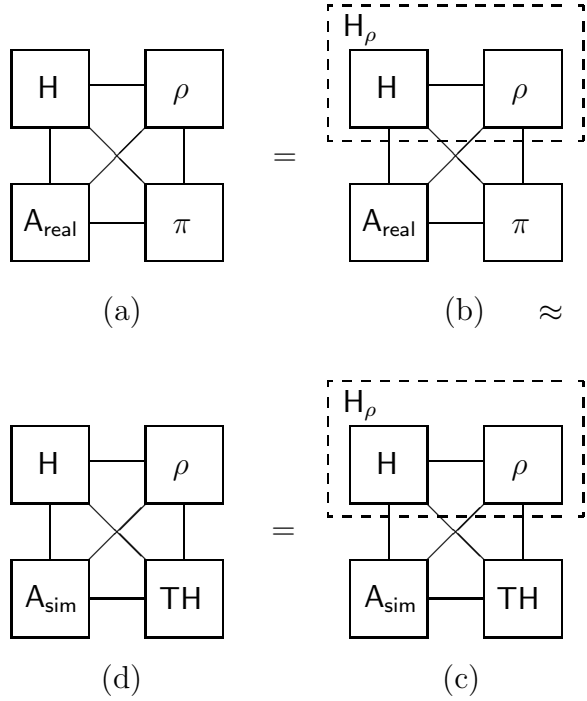

The basic idea of the proof of the simple composition theorem goes as
follows (we follow the proof idea from \cite{Pfitzmann:2001:Model},
but stay very general so that this proof idea should be applicable for
most simulatable security models):

Assume that $\pi\geq\sfTH$ and that $\rho^\sfTH\geq\mathsf{CPLX}$.
Consider any real adversary $\sfAr$ and honest user $\sfH$. The
network resulting from $\rho^\pi$, $\sfH$ and $\sfAr$ running together
is depicted in Figure~\ref{fig:proof.sc}a. Here we consider the
protocol $\rho^\pi$ to consist of machines executing $\rho$ and
machines executing $\pi$, the former connected to the latter by secure
connections.

Now consider a machine $\sfH_\rho$ simulating $\sfH$ and all the
machines in $\rho$. We can now replace $\sfH$ and $\rho$ by this
simulation and get the network depicted in Figure~\ref{fig:proof.sc}b.
Since $\sfH_\rho$ is a faithful simulation, the view of the original and of the simulated $\sfH$ are identical.

Note now that the network in Figure~\ref{fig:proof.sc}b is represents
the protocol $\pi$ running with honest user $\sfH_\rho$ and real
adversary $\sfAr$. Since we assumed $\pi\geq\sfTH$ we know, that there
is a simulator $\sfAs$ s.t.~the view of $\sfH_\rho$ is
indistinguishable in the networks of Figures~\ref{fig:proof.sc}b and
\ref{fig:proof.sc}c. From this we then conclude that the view of the
simulates $\sfH$ are also indistinguishable in these two networks.

Now we replace $\sfH_\rho$ again by the machines it simulated and get
the network in Figure~\ref{fig:proof.sc}d. Again, as $\sfH_\rho$ is
a faithful simulation, we now that the views of the simulated and the
original $\sfH$ in the networks in Figures~\ref{fig:proof.sc}c and
\ref{fig:proof.sc}d are identical.

We can then conclude that the view of $\sfH$ in the networks in
Figures~\ref{fig:proof.sc}a and \ref{fig:proof.sc}d are
indistinguishable, so we found a simulator $\sfAs$ for the composed
protocol $\rho^\pi$. Since we showed this for arbitrary $\sfH$ and
$\sfAr$, it follows $\rho^\pi\geq\rho^\sfTH$.

Using the transitivity of $\geq$,\footnote{The transitivity is obvious
  if the security notion is defined symmetrically, i.e.~if for the
  real and the ideal protocol the same adversaries are allowed and the
  scheduling etc.~are defined in the same for real and ideal
  execution. This is the case in the model of
  \cite{Backes:2004:Secure}, unfortunately not in the model of
  \cite{Canetti:2001:Security}, so that a formal proof in that model has
  to take care of more details.} from this and the assumption
$\rho^\sfTH\geq\mathsf{CPLX}$ we conclude $\rho^\pi\geq\mathsf{CPLX}$.
So the simple composition theorem is shown for the case of standard
security.

In the case of universal security, we additionally have to show that
the constructed simulator $\sfAs$ does not depend on the honest user
$\sfH$. However, since in Figure~\ref{fig:proof.sc}c $\sfAs$ does not
depend on $\sfH_\pi$, only on $\sfAr$, this follows trivially.

\subsection{Models of corruption}
\label{sec:corruption}
\index{corruption}

Another important point is the modelling of corruption. Since we
cannot assume all parties partaking in the protocol to be honest, we
have to assume that the adversary can corrupt some parties, which
afterwards are under his control.

We distinguish two main flavours of corruption: static and adaptive
corruption.

In the case of \emph{static
  corruption}\biindex{static}{corruption}\biindex{static}{security},
the adversary may choose a set of parties to corrupt (within some
limits, e.g.~at most $t$ parties) \emph{before} the protocol begins.
This models the idea parties are either dishonest or honest, but do
not change between these two states.  (Static corruption is the
default modelling in the model of \cite{Backes:2004:Secure}.)

In the case of \emph{adaptive
  corruption}\biindex{adaptive}{corruption}\biindex{adaptive}{security},
the adversary can at any time during the execution of the protocol
corrupt further machines (as long as the set of machines does not
exceed the given limits). In particular, the choice which party to
corrupt may depend on what is intercepted from a run of the protocol.
This model captures the view that an adversary may at any time try to
``persuade'' some parties to do his bidding, e.g.~by using force or
hacking into their computer.  (The model of
\cite{Canetti:2001:Security} captures this notion of corruption.)

It is common (both in \cite{Backes:2004:Secure} and
\cite{Canetti:2001:Security}) that the honest user/environment is
informed, which parties are corrupted. While it may seem strange at a
first glance, informing the environment has the following advantage:
Since the simulator tries to mimic the adversary exactly, the same
parties are corrupted in the ideal model. Assume now that we have
shown some protocol to be secure in the presence of at most $t+1$
corrupted parties. Then by using the fact that the simulator is
restricted to corrupting no more parties than the real adversary, we
can conclude that the protocol is also secure in the presence of at
most $t$ corrupted parties. This simple and quite intuitive result
would not hold if the real adversary and the simulator would not be
forced to corrupt the same number of parties.

When considering static corruption and requiring adversary and
simulator to corrupt the same parties, the modelling of a protocol
with corruptible parties can be reduced to that of a protocol with
only incorruptible ones using the following approach from
\cite{Backes:2004:Secure}: For each set $C$ of parties that may be
corrupted, let $\pi_C$ denote the protocol where these parties are
removed and all connections these parties had are instead connected to
the adversary. Let further $\sfTH_C$ denote the trusted host, where
all connections associated with the parties from $C$ are connected to
the adversary. Then $\pi$ implementing $\sfTH$ (with possible
corruption) is simply defined as $\pi_C\geq\sfTH_C$ (without
corruption) for all allowed sets $C$ of corrupted parties.

In the case of adaptive security, another important design choice
appears: whether parties are able to delete information. If they are
non-deleting, an adversary corrupting a party does not only learn all
information the party still wants to use in later protocol phases, but
also all information that ever came to the attention of that party.
This of course gives the adversary some additional advantage (but also
the simulator is given some advantage, so these two security with and
without non-deleting parties are probably incomparable). In
\cite{Canetti:2001:Security} non-deleting parties are used, while in
\cite{Backes:2004:Secure} deleting ones are used.

It is difficult to transport the notion of non-deleting parties to the
quantum case. Non-deleting parties cannot be required, since this
would imply cloning all information passing through the party. The
nearest analogue would seem to be non-measuring parties
\cite{Mueller-Quade:2002:PersonalSg}, which instead of measuring would
entangle the state to be measured with some ancillae. However, this
does not completely capture the notion of non-deleting parties, since
a party may now circumvent this restriction by basing some scheduling
related decisions on some bit.  If the scheduling is classical (see
Section~\ref{sec:sched}) this would destroy information.

Due to these problems in the modelling of non-deleting parties in the
quantum case, we will only model deleting parties in the sense that
parties are allowed to perform any measurements or erasures. Further
we will omit a discussion of adaptive corruption in the quantum case.

\subsection{Modelling machines}
\label{sec:machines}
\biindex{quantum}{machine}

Another point which can make subtle differences between different
simulatable security models, is the question how a machine is
modelled.

In classical models, the following two approaches are most common:
\begin{itemize}
\item In \cite{Backes:2004:Secure} a machine is modelled in a most
  general way. It is defined by a transition function that for each
  state of the machine, and each set of inputs gives a probability
  distribution over the output and state after that activation. Here
  the machine model does not a priori have the notion of a single
  computational step (like e.g.~a Turing machine would have), but only
  that of an activation. One can then however describe certain subsets
  of machines like the machines realisable by a Turing machine, or the
  machines realisable by a polynomial Turing machine etc. The
  advantage of this model is that one is not forced to formulate all
  arguments in terms of Turing machines, but in terms of the
  mathematically much easier transition functions.
\item In \cite{Canetti:2001:Security} a machine is generally assumed
  to be a polynomial interactive Turing machine. The non-deletion
  property is achieved by keeping a copy of the current state in each
  Turing step.
\end{itemize}

If adaptive corruption (see Section~\ref{sec:corruption}) is used,
another point must be taken care of: if a machine is corrupted, the
adversary learns its state, which means that there must be a canonical
interpretation how the state of a machine is encoded into a message
sent to the adversary.

In a quantum setting, one more choice has to be made:
\begin{itemize}
\item Should machines be forced to be unitary (and simulate
  measurements by entangling with auxiliary qubits), or should they be
  allowed to perform measurements.  At first it may seem, that unitary
  machines yield the simpler mathematical model. However, as soon as
  some kind of classical scheduling is involved (see
  Section~\ref{sec:sched}), measurements will take place anyway, so
  the overall behaviour of the network lacks a natural unitary
  description. When choosing machines to be allowed to perform
  measurements, then machines and network fall in a natural way into
  one mathematical framework. Further no options are lost by such a
  choice, since unitary transformations are a special case of 
  \emph{quantum superoperators} (see Section~\ref{sec:quant}).
\end{itemize}

In this work we have chosen to follow the approach of
\cite{Backes:2004:Secure} and try to achieve a modelling of machines
which captures any process not disallowed by physical laws.  This
yielded the following notion of machines:
\begin{itemize}
\item Following the \cite{Backes:2004:Secure}-approach, each machine
  has a set of in- and out-ports, modelling the connections to other
  machines. A port can be classical or quantum.
\item The transition of a machine is modelled by a quantum
  superoperator. It takes the state and inputs of the machine to the
  new state and outputs.
\item Before and after each activation, the classical ports are
  measured, the results of these measurements over the whole life of
  the machine constitute the view of the machine.
\item Before each activation a special (machine dependent) repeatable
  measurement is performed on the machine's state to decide, whether
  the machine is in a final state or not. This measurement may be
  trivial (for non-terminating machines).
\end{itemize}

\subsection{Computational security}
\label{sec:comp.sec}
\biindex{computational}{security}

The design choice which turns out to be the most difficult and
error-prone in definitions of simulatable security is how to model
computational security. 

The simplest and most common approach is to require all machines in a
network (adversary, simulator, honest user, protocol machines) to be
polynomially limited in the sense that there is a polynomial limiting
the overall number of Turing steps (or gates, etc.~depending on the
underlying model of computation). Even this seemingly simple approach
can lead to unexpected problems. Assume some trusted host is modelled
as a polynomial machine. Since reading a message needs time, the
machine will terminate after a given number of inputs on port $\mathsf
a$. Then however it will not be able to react to queries on port
$\mathsf b$, even if in the intuitive specification these ports are
completely unrelated. This is an artefact of the modelling and should
be avoided. Both \cite{Pfitzmann:2001:Model} and
\cite{Canetti:2001:Security} have these problems, the security proofs
in these models sometimes being strictly spoken incorrect, e.g.~most of the
functionalities in \cite{Canetti:2001:Security} were not polynomial
time, and therefore could not be composed (see below).

The obvious solution to allow real or ideal protocol to be
computationally unlimited and only restricting honest user, adversary
and simulator to be polynomial time, does not work either. An
inspection on the proof sketch of the simple composition theorem
(Section~\ref{sec:comp.flavours}) shows, that if the calling protocol
($\rho$ in the proof sketch) is not polynomially limited, then the
combination of $\rho$ and the honest user $\sfH$ is not polynomial
time any more. So the assumption $\pi\geq\sfTH$ cannot be used to
establish equivalence of the networks in Figures~\ref{fig:proof.sc}b
and \ref{fig:proof.sc}c and the proof fails. So such an approach to
computational security would lose the composition theorem and is thus
not viable.

In \cite{Backes:2002:Cryptographically} this problem was solved for
the RS framework by introducing so-called length functions. These
allowed a machine to selectively switch off ports, thus being able to
ignore inputs on these ports without losing computation time. This
solved the problem described above. To the best of our knowledge, in the UC
framework the problem has not yet been successfully approached.

However, the modelling using length functions still leaves some
problems. For example, it is not possible to model trusted hosts that
are able to process an arbitrary amount of data. An example for such a
trusted host would be a secure channel that is not a priori restricted
to some maximal number of messages. In the modelling using length
functions every trusted host (and every corresponding real protocol)
has to be parametrised in a ``life time polynomial'', which may be
arbitrarily chosen, but has to be fixed in advance.

A further possible solution is to allow parties to be polynomially
bounded in the number of activations or in the length of their inputs.
But this approach may lead to a modelling where two parties may gain
unbounded computation time by sending messages to each other in a kind
of ``ping-pong''\index{ping-pong} game.

A solution to this problem is presented in
\cite{Hofheinz:2004:Polynomial}. Using the modelling presented there
one can model parties that are able to process a not-a-priori-bounded
amount of data, without losing the property of being in an intuitive
way computationally bounded, and without losing the possibility of
composition. This additional generality is however bought by additional
complexity.

Other points that have to be detailed are e.g.~how much time is
consumed by reading a message, is a message read from the beginning
(disallowing to read the end of very long messages) or is there
random-access to the message content (i.e.~the machine can start
reading at the end, in the middle of a message, etc.\footnote{ In the
  quantum case even the access model would have to be distinguished,
  where a polynomial machine can in superposition access all bits of a
  message (the message is given to the machine as an oracle). Then the
  machine could e.g.~determine whether a message of exponential length
  is balanced or not \cite{Deutsch:1992:Rapid}, which clearly would
  not be possible if it would have to read the message
  symbol-by-symbol.}).

A noteworthy point is that (strict) statistical security does not
necessarily imply computational security. To see this, imagine an
ideal protocol that gives a computationally hard problem to the
simulator, and only if the simulator solves this problem, then the
ideal protocol will behave exactly as does the real protocol.  Then a
computationally unbounded simulator will be able to mimic the
behaviour of the real protocol, so we have strict statistical (even
perfect) security. But a computationally bounded simulator will be
unable to ``unlock'' the ideal protocol, so we do not have
computational security. A possible remedy to this problem has been
proposed by \cite{Canetti:2000:Security}: the simulator's runtime must
be polynomially bounded in the runtime bound of the real adversary.
Then statistical security implies computational security.

In the present work we concentrate on unconditional security.
However, we model the length functions of
\cite{Backes:2002:Cryptographically}, so computational security should
be easily definable following the example of
\cite{Backes:2004:Secure}.  The approach of
\cite{Hofheinz:2004:Polynomial} should be easily adaptable to the
model presented here, too.

\subsection{Scheduling and message delivery}
\label{sec:sched}
\index{scheduling}\biindex{message}{delivery}

Besides specifying how machines operate and send or receive messages,
it is necessary to model the behaviour of the overall network. It
turns out that here the details may get very complicated. We sketch
several general approaches to scheduling and message delivery:
\begin{itemize}
\item \emph{No scheduling.}\biindex{no}{scheduling} Here the order of messages and activations
  is determined in advance. For example, in a two party protocol it
  makes sense to say that the parties are alternatingly activated and
  upon each activation a message is sent to the other party. This is
  the simplest form of scheduling, it may however be unable to model
  the behaviour of more complex protocols where the decision whom to
  sent a message may depend on the protocol input or on some prior messages.
\item \emph{Message driven scheduling\biindex{message
      driven}{scheduling} with immediate message
    delivery\triindex{immediate}{message}{delivery}.}  Any machine can
  only send one message. This message is immediately delivered to the
  recipient (which may be the adversary in the case of an insecure
  channel).  This scheduling is very easy to model, but of course not
  every protocol can be described in such a model, and some realistic
  attacks may not be modelled. Further it seems difficult to model an
  authenticated channel (who should be activated next? the recipient?
  the adversary?).
\item \emph{Message driven scheduling with adversarially controlled
    delivery\triindex{adversarially controlled}{message}{delivery}.}
  Here a machine can send one or several messages in any activation.
  The adversary may then decide whether and when to deliver that
  message. Several flavours exist:
  \begin{itemize}
  \item \emph{Fair\triindex{fair}{message}{delivery} vs. asynchronous
      delivery\triindex{asynchronous}{message}{delivery}.} It has to be
    chosen, whether the adversary is required to eventually send a
    message (\emph{fair delivery}) or whether it may drop messages at
    will. In case of fair delivery, great care has to be taken with
    respect to the exact modelling, since otherwise the definition
    could e.g.~allow the adversary to ``deliver'' after the protocol's
    end.  Another question is whether parties may or may not know an
    upper bound for the time it takes a message to be delivered.  This
    problem is elaborated in more detail in \cite{Backes:2004:Fair}.
    Both \cite{Backes:2004:Secure} and \cite{Canetti:2001:Security}
    model asynchronous delivery. In \cite{Canetti:2001:Security}
    so-called \emph{non-blocking adversaries} are mentioned which are
    required to eventually deliver, but no definition is given and the
    above mentioned questions are not answered.

    Following \cite{Backes:2004:Secure}, we adopt asynchronous
    delivery, however the discussion from \cite{Backes:2004:Fair} can
    easily be adapted to our modelling.
  \item \emph{Blind\triindex{blind}{message}{delivery} or transparent
      delivery\triindex{transparent}{message}{delivery}.} Is the
    adversary notified on whether a message is to be delivered on some
    connection?  The adversary may have to schedule the connection
    without knowing whether some message is waiting (\emph{blind
      delivery}) or he is informed of the fact prior
    (\emph{transparent delivery}). The blind delivery captures the
    idea of a channel in which some reordering occurs, but which is
    not accessible to the adversary (e.g.~an ethernet connection). The
    transparent delivery captures the idea of a connection which is
    routed through adversarially controlled routers. In
    \cite{Backes:2004:Secure} blind delivery is used, while
    \cite{Canetti:2001:Security} adopts transparent delivery.

    We follow \cite{Backes:2004:Secure} and use blind delivery.
  \item \emph{Symmetric\triindex{symmetric approach
        to}{message}{delivery} or asymmetric
      approach\triindex{asymmetric approach to}{message}{delivery}.}
    In the \cite{Backes:2004:Secure} model each connection has a
    designated scheduler. These are usually the sending machine
    (modelling an immediately scheduled connection), the adversary (an
    adversarially controlled connection), or the recipient (a fetching
    connection).  We call this the \emph{symmetric
      approach}.\footnote{However, the symmetry is slightly broken by
      the fact that there is a designated machine called the
      \emph{master-scheduler}, which is activated if the activation
      token is lost.}  On the other hand \cite{Canetti:2001:Security}
    pursues an \emph{asymmetric approach}. Here many different rules
    of scheduling are defined depending on whether the machine is sent
    from functionality to parties or vice versa and environment and
    adversary have special uninterchangable roles.
    
    We adapt the symmetric approach from \cite{Backes:2004:Secure}
    since we believe that this makes the details of the modelling
    easier and is additionally of great advantage in detailed proofs,
    since much less different kinds of delivery are to be
    distinguished.
  \end{itemize}
\item \emph{Non-message driven scheduling\biindex{non-message
      driven}{scheduling}.} By this we mean the idea
  that all machines may run in parallel, and messages do not influence
  the scheduling. In particular, a machine may execute tasks without
  the need of being activated by an incoming message. A scheduling
  falling into that class has been described in
  \cite{Unruh:2002:Formal}. However, no satisfying and easy model has
  yet been defined using that approach.
  
  It is the author's personal opinion that such a model would capture
  reality much better, especially when allowing the machine to measure
  time in some way. However it seems that coming to an easy and
  intuitive modelling is a still unsolved problem.
\item \emph{Synchronous scheduling\biindex{synchronous}{scheduling}.} This means that the protocol
  proceeds in rounds, and in each round all machines are activated.
  This scheduling underlies e.g.~the early models
  \cite{Pfitzmann:2000:Secure} and \cite{Canetti:2000:Security}. However
  assuming synchronous scheduling means to assume a very strong
  synchronisation of the protocol participants and is only justified
  in special cases. This assumption was dropped in
  \cite{Pfitzmann:2001:Model} and \cite{Canetti:2001:Security}.
\end{itemize}

A further interesting issue is whether the scheduling and message
delivery is \emph{quantum}\biindex{quantum}{scheduling} or \emph{classical}\biindex{classical}{scheduling}. By quantum scheduling
we mean that events of scheduling (e.g.~the recipient of a message, or
the fact whether a message is sent at all, or which machine is
activated) can be in a quantum superposition.  In contrast, with
classical scheduling the state of the system would always collapse to
one of the possible decisions.

An advantage of quantum scheduling would be the possibility to model
protocols that explicitly make use of the superposition between
sending and not sending a message. For example, there is a protocol
that is able to detect if an eavesdropper tries to find out whether
communication takes place at all
\cite{Steinwandt:2002:Using,Mueller-Quade:2003:Problem} (traffic
analysis\biindex{traffic}{analysis}).

However, modelling quantum scheduling turns out to be quite difficult.
This is due to the fact that if the scheduling is to be non-measuring,
this means it has to be defined in an unitary way. But then it would
have to be reversible, thus disallowing many sensible machine
definitions. A suitable combination of measuring and non-measuring
scheduling would have to be found, capturing the possibilities of both
worlds.

Since how to model quantum scheduling is as far as we know an open
problem, we will here present a modelling of security using a
classical scheduling.

\subsection{The Ben-Or-Mayers model}
\label{sec:bom}

In this section we compare our model to the modelling of
\cite{Ben-Or:2004:General}. We organise this comparison into several
short topics, according to different sections of the introduction.

\begin{itemize}
\item \emph{Machine model (cf.~Section~\ref{sec:machines}).} In our
  modelling, machines are modelled by giving a transition operator,
  which can model any quantum-mechanically possible operation of the
  machine, including all types of measurements. In contrast, in the
  \cite{Ben-Or:2004:General}-model machines are modelled as partially
  ordered sets of gates\index{gate}, i.e., as circuits\index{circuit}
  (the partial order gives the order of execution of the gates).
\item \emph{The order of quantifiers
    (cf.~Section~\ref{sec:order.quant}).} In the
  \cite{Ben-Or:2004:General} model, no explicit real life adversary
  exists. Instead the environment communicates directly with the
  protocol, while in the ideal model a simulator is placed between the
  protocol and the environment (i.e., in the real model, a \emph{dummy
    adversary}\biindex{dummy}{adversary} is used). The idea behind
  such a dummy-adversary-approach (which is also given as an
  alternative formulation in \cite{Canetti:2001:Security}) is that if
  there are distinguishing adversary and environment, we could put the
  adversary into the environment and leave an empty hull in place of
  the adversary, the dummy adversary. Then the honest user would still
  distinguish. Therefore we can restrict our attention to such
  dummy-adversaries. Now, in the model of \cite{Ben-Or:2004:General},
  the order of quantifiers is as follows: For all environments, there
  is a simulator, such that real and ideal models are
  indistinguishable. So the simulator is chosen in dependence of the
  environment, so we have \emph{standard security} (while the notion
  of \emph{universal security} is not specified in that modelling).
\item \emph{Composition (cf.~Section~\ref{sec:comp.flavours}).} The
  \cite{Ben-Or:2004:General} shows a composition theorem which is
  comparable to what we call a \emph{simple composition theorem}.
  I.e., the concurrent composition of a non-constant number of copies
  of the protocols is not allowed by the composition theorem. Since
  the order of quantifiers is that of standard security, this seems
  natural.
  
  In the discussions in \cite{Ben-Or:2004:General}, some
  \emph{natural conditions for universal composition} are given, which
  seem to allow for concurrent composition of a polynomial number of
  protocol copies. However, these conditions require that the
  statistical distance between real and ideal protocol execution is
  bounded by a negligible function which must be \emph{independent of
    the environment in use.} This condition is a very strict one,
  since most computationally secure protocols violate that condition
  (consider some protocol where the adversary must find the pre-image
  of some function to break the protocol. Then the probability of
  breaking the protocol may go up if the adversary does a longer
  search for pre-images). In fact, the author knows of no protocol
  that is computationally but not statistically secure that would
  still be computationally secure with respect to these additional
  conditions.
\item \emph{Models of corruption (cf.~Section~\ref{sec:corruption}).}
  Corruption is modelled in the Ben-Or-Mayers model as follows: There
  are some so-called \emph{classical control
    registers}\triindex{classical}{control}{registers} which may control
  the order and application of the gates of a circuit. A corruptible
  machine will then execute its own gates if a special control
  register, the \emph{corruption register}, is set to $0$. If it is
  set to $1$, gates specified by the adversary are executed instead.
  Therefore in \cite{Ben-Or:2004:General} adaptive corruption is
  modelled, since the adversary might change the value of a control
  register at a later time.
\item \emph{Computational security (cf.~Section~\ref{sec:comp.sec}).}
  In \cite{Ben-Or:2004:General} computational security is modelled by
  requiring that for any polynomial $f$, any family of environments
  with less than $f(k)$ gates (where $k$ is the security parameter)
  achieves only negligible statistical distance between ideal and real
  protocol execution. This is roughly equivalent to requiring security
  against non-uniform environments, adversaries and simulators.
\item \emph{Scheduling and message delivery
    (cf.~Section~\ref{sec:sched}).}  We have tried to realise a very
  general scheduling using the concept of buffers, which are scheduled
  at the discretion of the adversary. In contrast, in the
  \cite{Ben-Or:2004:General}-model a more simple scheduling was
  chosen: The delivery of messages and activation of machines is
  represented by the ordering of the gates. Since the ordering of
  gates within one machine is fixed up to commuting gates, and the
  access of gates of different machines to a shared register (a
  \emph{channel register}) is defined to alternate between sender and
  recipient, all scheduling and message delivery between machines is
  fixed in advance. Only when the adversary gates are executed, the
  scheduling may have some variability. Further there might be
  situations (like a channel shared by more than two parties) where
  the scheduling is not completely fixed.
  
  However, the partial order on the gates is fixed by the environment
  \emph{in advance}, i.e., may not depend of information gathered by
  the adversary in the course of a protocol execution.
  
  This model of scheduling seems adequate for protocols with a very
  simple communication structure (like two-party protocols proceeding
  in rounds), more complex protocols however tend to contain the
  possibility of race conditions\index{race condition}\footnote{I.e.,
    the protocol is sensitive to the order of events. A typical race
    condition would be if some event happens exactly between two
    delicate steps of the protocol, causing confusion and insecurity.}
  and message reordering\biindex{message}{reordering}. Therefore a
  security guarantee given in the \cite{Ben-Or:2004:General}-model for
  such a protocol should be taken with care, since attacks based on
  such scheduling issues would not be taken into account by that model
  (at least, if the attack is based on information gathered by the
  adversary).
  
  Further it does not seem possible even to model protocols which contain
  instructions like ``toss a coin; on outcome $0$ send a message to
  Alice, on outcome $1$ send a message to Bob'', since the sending of
  messages is fixed in advance.
\item \emph{Service ports (cf.~Section~\ref{sec:protocols}).} Often it is
  useful to dedicate some in- and outgoing connection of the protocol
  to be used only by the adversary (in particular, the simulator can
  lie to the environment about the information on these connections).
  Other connections again should be used by the environment (since
  they constitute the protocol in- and outputs). We have realised this
  concept by specifying a set of protocol ports that can be used by
  the honest user, the \emph{service ports}. All other ports are
  restricted to be used only by the adversary.
  
  Such a distinction is particularly useful when specifying a trusted
  host, where some side-channels (like the size of a transmitted
  message or similar) are intended only for the adversary.
  
  It is unclear to us, whether the \cite{Ben-Or:2004:General} model
  provides a possibility for specifying such a distinction (i.e., for
  telling the simulator which ports it may access and which it may
  not).
\end{itemize}

It would be very interesting to study, how security in the
\cite{Ben-Or:2004:General}-model relates to security in our model.
E.g., a theorem like ``if the protocol $\pi$ is secure in
\cite{Ben-Or:2004:General} and satisfies such-and-such conditions,
then the protocol $\pi'$, resulting from translating $\pi$ into our
model in such-and-such a way, is secure in our model'' would seem very
nice. (The other direction might be more difficult, since many
protocols in our model that have more than three parties would not
have an intuitive counterpart in the
\cite{Ben-Or:2004:General}-model.)

\subsection{Quantum mechanical formalism used in this work}
\label{sec:quant}
\biindex{quantum mechanical}{formalism}

In this work, we adopt the formalism that states in quantum mechanics
can be described using \emph{density
  operators}\biindex{density}{operator}. Given a \emph{Hilbert
  space}\biindex{Hilbert}{space} $\calH$, the set $\Rho(\calH)$ of
positive linear operators on $\calH$ with trace $1$.

If $\calH=\setC^X$ for some countable set $X$, we say that $\{\ket
x:x\in X\}$ is the \emph{computational
  basis}\biindex{computational}{basis} for $\calH$.

Any operation on a system can then described by a so-called
\emph{superoperator}\index{superoperator}\index{operator!super-} (or
\emph{quantum operation}\biindex{quantum}{operation}).  A mapping
$\calE:\Rho(\calH)\rightarrow\Rho(\calH)$ is a superoperator on
$\Rho(\calH)$ (or short on $\calH$), iff it is convex-linear,
completely positive and it is $0\leq\tr\calE(\rho)\leq\tr\rho$ for all
$\rho\in\Rho(\calH)$.

Superoperators $\calE$ on a system $\calH_1$ can be extended to a
larger system $\calH_1\otimes\calH_2$ by using the tensor product
$\calE\otimes 1$ (where $1$ is the identity).

We will only use a special kind of measurement, so-called
\emph{von-Neumann}\biindex{von-Neumann}{measurement} or \emph{projective measurements}\biindex{projective}{measurement}. A von-Neumann
measurement on $\calH$ is given by a set of projections $P_i$ on
$\calH$, s.t.~$\sum_i P_i=1$ and all $P_i$ are pairwise orthogonal,
i.e.~$P_iP_j=0$ for $i\neq j$. Given a density operator
$\rho\in\Rho(\calH)$, the probability of measurement outcome $i$ is
$\tr P_i\rho$, and the post-measurement state for that outcome is
$P_i\rho P_i$.

We say a von-Neumann measurement is complete, if every projector has
rank $1$. A complete von-Neumann measurement in the computational
basis denotes a von-Neumann measurement where each projector project
onto one vector from the computational basis of $\calH$.

Von-Neumann measurements too can be extended to a larger system by
using the tensor-product.

For further reading we recommend the textbook
\cite{Nielsen:2000:Quantum}.

Additionally we fix that $\setN$ denotes the natural number excluding
$0$, $\setN_0$ the natural number including zero, $\setR$ the real
numbers, and $\setR_{\geq0}$ the non-negative real numbers.

%%% Local Variables:
%%% mode: latex
%%% coding: latin-1
%%% TeX-master: "quantum-security"
%%% End:
%%% Local IspellDict: british

% LocalWords:  simulatability composable Backes Pfitzmann BPW simulatable

\section{Quantum networks with scheduling}
\label{sec:qns}

In this following we will sketch how networks of quantum machines can
be defined, while trying to closely mimic the behaviour of the
\emph{classical} networks of the RS framework of
\cite{Backes:2004:Secure}. 

Roughly, a quantum network (\emph{collection}\index{collection} in the
terminology of \cite{Backes:2004:Secure}) consists of a set of
machines. Each machine has a set of \emph{ports}, which can be
connected to other machines to transmit and receive messages. A port
can be of the following kinds:
\begin{itemize}
\item A simple out-port\biindex{simple}{out-port}\index{port!simple
    out-} $\port p!$. This is a port on which a machine can send a
  message.
\item A simple in-port\biindex{simple}{in-port}\index{port!simple
    in-} $\port p?$. This is a port on which a machine
  can receive a message.
\item A clock out-port\biindex{clock}{out-port}\index{port!clock
    out-} $\port p<!$. Using this port a machine can
  schedule a message, i.e.~messages sent from port $\port p!$ are not
  delivered to $\port p?$ until they are scheduled via $\port p<!$.
\item Clock in-ports\biindex{clock}{in-port}\index{port!clock in-}
  $\port p-?$, buffer
  in-ports\biindex{buffer}{in-port}\index{port!buffer in-} $\port p-?$
  and buffer out-ports\biindex{buffer}{out-port}\index{port!buffer
    out-} $\port p-!$. These ports are special to the so-called
  \emph{buffers} and are explained below.
\item The
  master-clock-port\index{master-clock-port}\index{port!master-clock}\index{clock
    port}{master-} $\masterclk$. This special case of a clock in-port
  is special in that it is not connected to another machine.  Instead,
  a special machine called the \emph{master scheduler} is activated
  via this port if the activation token is lost, i.e.~if no machine
  would be activated via an incoming message.
\end{itemize}

Each port is additionally classified as being either a \emph{quantum}
or a \emph{classical port}. (This of course is an addition to the
modelling of \cite{Backes:2004:Secure}.)

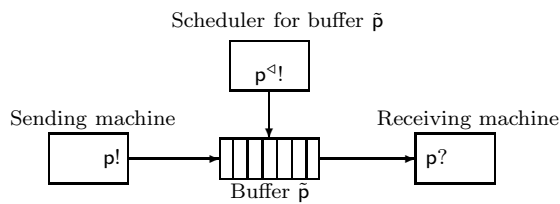
\begin{figure}[thbp]
  \begin{center}
    \setlength{\unitlength}{2693sp}%
\begingroup\makeatletter\ifx\SetFigFont\undefined%
\gdef\SetFigFont#1#2#3#4#5{%
  \reset@font\fontsize{#1}{#2pt}%
  \fontfamily{#3}\fontseries{#4}\fontshape{#5}%
  \selectfont}%
\fi\endgroup%
\begin{picture}(4467,1773)(1216,-2725)
\thinlines
{\color[rgb]{0,0,0}\put(4951,-2536){\framebox(720,450){}}
}%
\put(5041,-2356){\makebox(0,0)[lb]{\smash{\SetFigFont{8}{9.6}{\familydefault}{\mddefault}{\updefault}{\color[rgb]{0,0,0}${\sf p}?$}%
}}}
{\color[rgb]{0,0,0}\put(3267,-2131){\line( 0,-1){360}}
}%
{\color[rgb]{0,0,0}\put(3402,-2131){\line( 0,-1){360}}
}%
{\color[rgb]{0,0,0}\put(3537,-2131){\line( 0,-1){360}}
}%
{\color[rgb]{0,0,0}\put(3672,-2131){\line( 0,-1){360}}
}%
{\color[rgb]{0,0,0}\put(3807,-2131){\line( 0,-1){360}}
}%
{\color[rgb]{0,0,0}\put(3942,-2131){\line( 0,-1){360}}
}%
{\color[rgb]{0,0,0}\put(2296,-2311){\vector( 1, 0){855}}
}%
{\color[rgb]{0,0,0}\put(1576,-2536){\framebox(720,450){}}
}%
{\color[rgb]{0,0,0}\put(4051,-2311){\vector( 1, 0){900}}
}%
{\color[rgb]{0,0,0}\put(3151,-2491){\framebox(900,360){}}
}%
{\color[rgb]{0,0,0}\put(3601,-1681){\vector( 0,-1){450}}
}%
{\color[rgb]{0,0,0}\put(3241,-1681){\framebox(720,450){}}
}%
\put(1216,-1996){\makebox(0,0)[lb]{\smash{\SetFigFont{8}{9.6}{\familydefault}{\mddefault}{\updefault}{\color[rgb]{0,0,0}Sending machine}%
}}}
\put(4591,-1996){\makebox(0,0)[lb]{\smash{\SetFigFont{8}{9.6}{\familydefault}{\mddefault}{\updefault}{\color[rgb]{0,0,0}Receiving machine}%
}}}
\put(3241,-2671){\makebox(0,0)[lb]{\smash{\SetFigFont{8}{9.6}{\familydefault}{\mddefault}{\updefault}{\color[rgb]{0,0,0}Buffer $\tilde{\sf p}$}%
}}}
\put(2071,-2356){\makebox(0,0)[lb]{\smash{\SetFigFont{8}{9.6}{\familydefault}{\mddefault}{\updefault}{\color[rgb]{0,0,0}${\sf p}!$}%
}}}
\put(2701,-1096){\makebox(0,0)[lb]{\smash{\SetFigFont{8}{9.6}{\familydefault}{\mddefault}{\updefault}{\color[rgb]{0,0,0}Scheduler for buffer $\tilde{\sf p}$}%
}}}
\put(3466,-1591){\makebox(0,0)[lb]{\smash{\SetFigFont{8}{9.6}{\familydefault}{\mddefault}{\updefault}{\color[rgb]{0,0,0}${\sf p}^{\triangleleft}!$}%
}}}
\end{picture}
  \end{center}
  \vskip-5mm
  \caption{A connection}
  \label{fig:buffer}
\end{figure}

There are further special machines called buffers\index{buffer}. These
have the task of storing messages while they are waiting to be
delivered. A buffer $\buff p$ necessarily has ports $\port p-?$,
$\port p-!$ (buffer in-/out-port) and $\port p<?$ (clock in-port). It
is further associated to the ports $\port p!$, $\port p?$ and $\port
p-!$ by its name $\mathsf p$. Therefore we get the situation depicted
in Figure~\ref{fig:buffer}.\index{connection}

A message delivery now takes place at follows: The sending machine
sends a message by writing a string (preparing a quantum state) on the
out-port $\port p!$. This message is then immediately transfered into
the buffer. In the buffer it appended to a queue. In this queue it
stays until the scheduler for buffer $\buff p$ sends a natural number
$n$ on $\port p<!$. If $n\geq1$ and there are at least $n$ message in
the buffer's queue, the $n$-th message is removed from the queue and
transmitted to the simple in-port $\port p?$ of the receiving machine.
The receiving machine is then activated next.

The three machines in Figure~\ref{fig:buffer} are not necessarily
different ones. In particular, it may e.g.~be that the sending machine
schedules its messages itself, realising immediate delivery, or that a
machine sends messages to itself.

Note that if a machine has several clock out-ports and sends a number
on several of these, all but the first one (in some canonical ordering
of the ports) are ignored, since otherwise it would be unclear which
machine to activate next.

If no machine would be activated next (e.g., because the machine last
activated did not write on a clock out-port, or because the number $n$
sent through the clock out-port $\port p<!$ was higher than the number
of messages queued in the buffer $\buff p$) a special designated
machine is activated, the \emph{master
  scheduler}\biindex{master}{scheduler}. This machine is characterised
by having the master-clock-port $\masterclk$ on which it then gets the
constant input $\mathsf 1$.

We have now seen that in a network there are three types of machines
\begin{itemize}
\item A \emph{simple machine}\biindex{simple}{machine}. This machine is
  characterised by having only simple in- and out-ports and clock
  out-ports.
\item A \emph{master scheduler}. This machines may have the same ports
  as a simple machine, but additionally has the master-clock-port
  $\masterclk$. Also the master scheduler is the machine to be
  activated at the beginning of the execution of the network.
\item A \emph{buffer}. This machine which has a completely fixed
  behaviour exists merely to store and forward messages. (The buffer
  will be more exactly defined below.)
\end{itemize}

Upon each activation of some simple machine or master scheduler, a
record of this activation is added to the so-called
\emph{trace}\index{trace} or \emph{run}\index{run} of the network (or
collection). This record consists of the name of the machine, of all
its \emph{classical} inputs (contents of the classical in-ports before
activation), all its \emph{classical outputs} (contents of the
classical out-ports after activation), and the \emph{classical state}
(see below) before and after activation.\footnote{Here we slightly
  deviate from the modelling of \cite{Backes:2004:Secure}, where the
  complete in-/output and the complete state is logged.  This of
  course is not possible in the quantum case since it would imply
  measuring the quantum state and all the in-/outputs in every
  activation.}

From the run we can easily extract the so-called \emph{view} of some
machine $\mathsf M$. It consists of all records in the run containing
the name of $\mathsf M$.

The scheduling as described above will be transformed into a formal
definition (while strongly drawing from the definitions from
\cite{Backes:2004:Secure} where the introduction of quantum mechanics
does not necessitate an alteration).

\subsection{Quantum machines}
\label{sec:mach}
\biindex{quantum}{machine}
  
First, for self-containment, we restate the unchanged formal
definition of a port from \cite{Backes:2004:Secure}:

\begin{definition}[Ports \cite{Backes:2004:Secure}]\index{port}
  Let $\mathcal
  P:=\Sigma^+\times\{\varepsilon,{}^\leftrightarrow,{}^\triangleleft\}\times\{!,?\}$
  (here $\varepsilon$ denotes the empty word). Then $p\in\mathcal P$
  is called a \emph{port}. For $p=(\mathsf n,l,d)\in\calP$, we call
  $\mathsf{name}(p):=\mathsf n$ its \emph{name}, $\mathsf{label}(p):=l$ its
  label, and $\mathsf{dir}(p):=d$ its direction.
\end{definition}

Usually we do not write $(\mathsf n,l,d)$, but $\mathsf nld$, i.e.,~a
port named $\mathsf p$ with label ${}^\triangleleft\}$ and direction
$!$ would simple be written $\port p<!$, if the label was
$\varepsilon$, we would write $\port p!$.

Note that label $\varepsilon$ denotes a simple port, label
${}^\leftrightarrow$ a buffer port and label ${}^\triangleleft$ a
clock port. Further direction $!$ denotes an out-, and direction $?$ an
in-port.

Further, if $P$ is a set or sequence of ports, let $\mathsf{in}(P)$
and $\mathsf{out}(P)$ denote the restriction of $P$ to its in- or
out-ports, resp.

We can now proceed to the definition of a machine. Since our model
shall encompass quantum machines, we will here deviate from the
modelling of machines in \cite{Backes:2004:Secure}. However, to
simplify comparisons, we shortly recapitulate the definition of a
machine in \cite{Backes:2004:Secure} (which we will call a
BPW-machine\index{machine!BPW-}\index{BPW-machine}):
\begin{itemize}
\item A machine $\mathsf M$ is a tupel
  $M=(\textit{name},\textit{Ports},\textit{States},\delta,l,\textit{Ini},\textit{Fin})$.
  Here $\textit{name}$ is the unique name of the machine,
  $\textit{Ports}$ the sequence of the ports of this machine,
  $\textit{States}\subseteq\Sigma^*$ the set of its possible states.
\item $l$ is the length function of this machine, see below.
\item $\mathit{Ini}$ is the set of initial states\biindex{initial}{state}. Since in the
  security definitions in \cite{Backes:2004:Secure} only the states of
  the form $\mathsf 1^k$ are used (where $k$ is the security
  parameter), we can w.l.o.g.~assume $\mathit{Ini}=\{\mathsf
  1^k:k\in\setN\}$.
\item $\textit{Fin}$ is the set of final states. A machine reaching a state in
  $\textit{Fin}$ will never be activated again.
\item $\delta$ is the \emph{state-transition function}\biindex{state-transition}{function}. For a given
  state $s$ of the machine, and inputs $I$, $\delta(s,I)$ gives the
  probability distribution of $(s',O)$, where $s'$ is the state of the
  machine after activation, and $O$ its output.
\end{itemize}

In comparison, we define a machine in our quantum setting as follows
(cf.~also the discussion after this definition):
\begin{definition}[Machine]
  A \emph{machine} (or \emph{quantum
    machine}\biindex{quantum}{machine}) is a tuple $$
  M =
  (\textit{name},\textit{Ports},\textit{CPorts},
  \textit{QStates},\textit{CStates}, \Delta,l,\textit{Fin}) $$
  where
  \begin{itemize}
  \item $\textit{name}\in\Sigma^+$ is the name of the
    machine\index{name of a machine}.
  \item $\textit{Ports}$ is the sequence of the ports of the machine.
  \item $\textit{CPorts}\subseteq\textit{Ports}$ is the set of the
    classical ports\biindex{classical}{port}\biindex{quantum}{port} of
    the machine. There must be no clock-ports in
    $\textit{Ports}\setminus\textit{CPorts}$ (i.e.~all clock-ports are
    classical).
  \item $\textit{QStates}\subseteq\Sigma^*$ is the basis of the space
    of the quantum states\biindex{quantum}{state}, i.e.~the states of $\mathsf M$ live in
    $\setC^{\textit{QStates}}$. It must be $\mathsf
    0\in\mathit{QStates}$.
  \item $\textit{CStates}\subseteq\Sigma^*$ is the set of the
    classical states\biindex{classical}{state} of the machine. It must be $\mathsf
    1^k\in\textit{CStates}$ for all $k\in\setN$ (i.e.~the classical
    states must be able to encode the initial states).
  \item The \emph{state-transition
      operator}\biindex{state-transition}{operator} $\Delta_M$ is a
    trace-preserving superoperator operating over the Hilbert space
    $\setC^{\textit{QStates}}\otimes\setC^{\textit{CStates}}
    \otimes\setC^\calI\otimes\setC^\calO$.  Here
    $\calI:=(\Sigma^*)^{\mathsf{in}(\textit{Ports})}$ is the set of
    all possible inputs of $\mathsf M$ (strictly spoken the basis of
    the state of all inputs), and
    $\calO:=(\Sigma^*)^{\mathsf{out}(\mathit{Ports})}$ analogously for
    the outputs of $\mathsf M$.
  \item The function
    $l:\mathit{CStates}\times\mathsf{in}(\Ports)\rightarrow\{0,\infty\}$
    is called the \emph{length function}\biindex{length}{function} of $\mathsf M$.  For each
    classical state $c$ and each in-port $\port p$, this tells whether
    input on this port should be ignored ($l(c,\port p)=0$) or not
    ($l(c,\port p)=\infty$).
  \item $\mathit{Fin}\subseteq\mathit{CStates}$ is the set of final
    states\biindex{final}{state}. If the classical state of the machine is a final state,
    then the machine will not be activated any more.
  \end{itemize}
  
  Given a machine $\mathsf M$, we denote the different entries of the
  tupel defining $\mathsf M$ by the name of the entry and a subscript
  $\mathsf M$. E.g., $\mathit{CStates}_\mathsf M$ are the classical states of
  $\mathsf M$.
\end{definition}

We will now discuss the elements of this definition. The field
$\mathit{name}$ simply defines a unique name of the machine which is
used to know with which machine the entries in the trace are associated.

The sequence $\mathit{Ports}$ denotes, which ports a given machine
has. A subset of these are the classical ports $\mathit{CPorts}$. All
messages written to or read from the classical ports are measured in
the computational basis (before or after the activation of $\mathsf
M$, depending on whether it is an in- or an out-port). Note that the
machine definition above does not handle classical ports differently
from quantum ports, the measuring will take place in the run-algorithm
(see Section~\ref{sec:run}).  We do not allow clock-ports to be
quantum ports, since the clock ports contain numbers of the are to be
scheduled. Since our scheduling is classical
(cf.~Section~\ref{sec:sched}), these numbers have to be classical.

One notices, that the above machine definition has two sets of states,
the quantum states $\mathit{QStates}$ and the classical states
$\mathit{CStates}$. We may imagine the machine to be a bipartite
system, consisting of a classical part and a quantum part. However,
since the quantum part not restricted to unitary operations, this does
not imply a strict separation of a controlling classical machine and a
controlled pure quantum machine. This distinction is necessary, since
some events in the scheduling etc.~depend on the state of the machine.
Since these events are of classical nature (see discussion in
Section~\ref{sec:sched}), they may not depend on the quantum state.
Note again, that the state-transition operator does not treat the
classical and the quantum state differently, the run-algorithm
(Section~\ref{sec:run}) takes care of measuring the classical state.

The most interesting part of the machine is probably the
state-transition operator, since it specifies the behaviour of the
machine. We try to make the machine definition as general as possible,
therefore we want to allow any quantum mechanically possible operation
of the space accessible to the machine. This state consists of the
inputs $\setC^{\calI}$, the outputs $\setC^\calO$, and of course of
the bipartite state
$\setC^{\mathit{QStates}}\otimes\setC^{\mathit{CStates}}$ of the
machine. The most general operation on such a space is described by a
trace-preserving superoperator
(cf.~Section~\ref{sec:quant}).\footnote{We do, of course, neglect
  advanced physics like special and general relativity. However,
  formally modelling these in a security model is probably still far
  off future.} Therefore we do not impose any more restrictions on the
state-transition operator $\Delta$ than to be such a superoperator.
Note that $\Delta$ is formally even allowed to read its output space
or write to its input space. However, since the input space is erased
after activation, and the output space is initialised to a known state
before activation (cf.~Section~\ref{sec:run}), this does not pose a problem.

The \cite{Backes:2004:Secure} model introduced so-called \emph{length
  functions} to cope with some problems occurring when modelling
computational security (see~\ref{sec:run}). These length functions
allow machines to set the maximal length of a message which can be
received on a given in-port (longer input is truncated). So in
particular, a machine can switch off some in-port completely, which
has the effect that this machine is not activated any more on input on
that port. Since whether the machine is activated or not is here
defined as a classical decision, the length function should be
classically defined, too. Therefore the length function must only
depend on the classical state of the machine. In our setting the
length function only plays an inferior role, since we are only
concerned with unconditional security. However, since this model is
designed to be easily extendable to the computational case (by
defining a computational model for the machines and then using a
straightforward adaption of the definition of computational security
from \cite{Backes:2004:Secure}), we included the length functions into
our model. The only important values of length functions are $0$
(ignore the message), and $\infty$ (do not ignore it), all other
values (integers greater $0$) just truncate the message, but even a
computationally limited machine could simply ignore anything longer
than indicated by the length function. Since further truncating would
imply at least a partial measurement of the length of the message, we
have restricted the length functions to take only the values $0$ and
$\infty$.

Finally the decision whether a machine has terminated or not should be
classical, this should only depend on the classical state of the
machine. Therefore the set $\mathit{Fin}$ of final states is a subset
of $\mathit{CStates}$.

A special kind of machine is the so called buffer (see the informal
description in Section~\ref{sec:qns}). The state of the buffer
contains a (possibly empty) queue of messages. Note that the buffer
does not measure these messages, they are simply moved. A buffer
$\buff p$ can be called due to two different reasons:
\begin{itemize}
\item A message arrived on its buffer in-port $\port p-?$. Then this
  message is appended to the queue.
\item A number $n$ was written to its clock in-port $\port p<?$. Then
  the $n$-th message is taken from the queue (if existent) and moved
  to the buffer out-port $\port p-!$.
\end{itemize}
For completeness we give a formal definition of buffers:
\begin{definition}[Buffers]\label{def:buffer}\index{buffer}
  Let $\mathit{Queue}$ denote the set of all possible queue contents:
  $$
  \mathit{Queue} := \{(n;m_1,\dots,m_n) : n\in\setN_0, m_i\in\Sigma^*\}
  $$
  We assume the elements of $\mathit{Queue}$ to be encoded as words
  in $\Sigma^*$, so that $(0)$ (the empty queue) is encoded as the
  empty word $\varepsilon\in\Sigma^*$.

  The \emph{buffer} $\buff p$ is defined by
  $$
  \buff p := (\mathsf{n\mathord\sim},(\port p-?,\port p-!,\port
  p<?),\{\port p<!\},\mathit{Queue},\{\mathsf 1^k\},
  \Deltabuff,\infty,\emptyset)
  $$
  That is, the buffer is named $\mathsf{n\mathord\sim}$, has ports $\port
  p-?,\port p-!,\port p<?$, of which $\port p<!$ is classical, the
  classical states are the required initial states $\mathsf 1^k$, the
  quantum states are the possible queue contents (where the queue
  initially is empty). The length function is set to constant
  $\infty$, i.e.~no truncating takes place. The set of final states is
  $\emptyset$, so the buffer will never terminate.
  
  The state-transition operator $\Deltabuff$ is defined by the
  following measurement process on the buffer's state (in $
  \calH_{\buff p}= \setC^{\textit{Queue}}\otimes \setC^{\{\mathsf
    1^k\}}\otimes \calH_{\port p-?}\otimes\calH_{\port
    p<?}\otimes\calH_{\port p-!}$):
  \begin{itemize}
  \item Measure whether $\calH_{\port p-?}$ contains state
    $\ket\varepsilon$ ($\varepsilon$ being the empty word). If no
    ($\port p-?$ is nonempty), perform the linear operation given by
    $$
    \ket{(n;m_1,\dots,m_n)}\otimes\ket{\mathit{in}}
    \longmapsto
    \ket{(n+1;m_1,\dots,m_n,\mathit{in})}\otimes\ket{\varepsilon}
    $$
    on $\setC^{\textit{Queue}}\otimes\calH_{\port p-?}$. (I.e., if
    there is input on port $\port p-?$, append it to the queue.)
  \item Perform a complete von-Neumann measurement in the
    computational basis (from now on called complete measurement) on
    $\calH_{\port p<?}$. Let $i$ be the outcome.
  \item Prepare state $\ket\varepsilon$ in subsystem $\calH_{\port
      p-!}$.
  \item Measure the first component in $\setC^{\textit{Queue}}$
    (i.e.~project $\setC^{\textit{Queue}}$ onto one of the spaces
    $S_n$ where $S_n$ is spanned by the vectors
    $\ket{(n,m_1,\dots,m_n)}$, $m_i\in\Sigma^*$). Let $n$ be the
    outcome of this measurement (i.e.~$n$ is the current queue
    length).
  \item If $i\in\setN\subseteq\Sigma^*$ (we assume natural numbers to be
    encoded as nonempty words in $\Sigma^*$) and $i\leq n$, then perform the
    following linear operation on
    $\setC^{\textit{Queue}}\otimes\calH_{\port p-!}$:
    $$
    \ket{(n;m_1,\dots,m_n)}\otimes\ket{\varepsilon}
    \longmapsto
    \ket{(n;m_1,\dots,m_{i-1},m_{i+1},\dots,m_n)}\otimes\ket{m_i}.
    $$
    That is, the $i$-th message is moved to the buffer out-port
    $\port p-!$.
  \end{itemize}
\end{definition}

\subsection{Quantum networks}
\label{sec:run}
\biindex{quantum}{network}

So far we have only modelled single machines. From here, the
definition of a network is not very far away. In the
\cite{Backes:2004:Secure} modelling a network simply is a set of
machines (called a \emph{collection}). The connections between the
machines are given by the names of the ports, as described in
Section~\ref{sec:qns} and depicted in Figure~\ref{fig:buffer}. Some
restriction have to apply to a collection to form a sensible network,
e.g.~there must not be any dangling connection (free ports), and no
ports must be duplicated. The formal definition of collections is
literally identical to that in \cite{Backes:2004:Secure}, we cite it
for selfcontainedness:

\begin{definition}[Collections \cite{Backes:2004:Secure}]\index{collection}
  \begin{itemize}
  \item A collection $\Hat C$ is a finite set of machines with
    pairwise different machine names, pairwise disjoint port sets, and
    where each machine is a simple machine, a master scheduler, or a
    buffer.
  \item $\mathsf{ports}(\Hat C)$ denotes the set of all ports of all
    machines (including buffers) in $\Hat C$.
  \item If $\buff n$ is a buffer, $\buff n,\mathsf M\in\Hat C$ and
    $\port n<!\in\mathit{Ports}_\mathsf M$ then we call $\mathsf M$
    the \emph{scheduler} for buffer $\buff n$ in $\Hat C$, and we omit
    ``in $\Hat C$'' if it is clear from the context.
  \end{itemize}
\end{definition}

Note that a collection is not necessarily a complete network, since it
is not required that each port has its counterpart. This is important
since we will need these ``non-closed'' collections for defining the
notion of protocols (Section~\ref{sec:secdef}). 

Prior to introducing the notion of a \emph{closed collection}, which
will represent quantum networks, we have to introduce the notion of
the free ports. The \emph{low-level complement}\biindex{low-level}{complement} $\port p^c$ of some port
$\port p$ is the port which in Figure~\ref{fig:buffer} is directly
connected to $\port p$. That is, $(\port p!)^c=\port p-?$, $(\port
p?)^c=\port p-!$, $(\port p<!)^c=\port p<?$, $(\port p-?)^c=\port p!$,
$(\port p-!)^c=\port p?$, and $(\port p<?)=\port p<!$. Then we can
define the set of \emph{free ports}\biindex{free}{port} of a collection $\Hat C$ to be the set of
ports in $\Hat C$ that do not have a low-level complement in $\Hat C$,
formally $\mathsf{free}(\Hat C)=\{\port p\in\mathsf{ports}(\Hat
C):\port p^c\notin\mathsf{ports}(\Hat C)$. Intuitively, the free ports
are those that have a ``dangling'' connection.

We can now define a closed collection:

\begin{definition}[Closed collection, completion
  \cite{Backes:2004:Secure}]\biindex{closed}{collection}\biindex{completion of a}{collection}
\begin{itemize}
\item The completion $[\Hat C]$ of $\Hat C$ is defined as
  $$
  [\Hat C]:=\{\buff n\ \vert\ \exists l,d: (\mathsf n,l,d)\in\mathsf{ports}(\Hat C)\setminus\{\masterclk\}\}.
  $$
\item $\Hat C$ is closed iff $\mathsf{free}([\Hat C])=\{\masterclk\}$.
\end{itemize}
\end{definition}

Intuitively, the completion of $\Hat C$ results from adding to $\Hat
C$ all missing buffers. That is, if some machine has a port $\port
p!$, $\port p?$, or $\port p<!$, then the buffer $\buff p$ is added if
not yet present. Note that no buffer is added for the
master-clock-ports $\masterclk$, since this port should be left free
(it is only used to activate the master scheduler in case of a lost
activation token).

A collection is then called closed, if only buffers are missing,
i.e.~if after completing it, there is no ``dangling connection'' (note
that of course the master-clock-port must stay unconnected). Such a
closed collection is now a quantum network ready to be executed.

We will now proceed to defining the run of a network. In contrast to
the definitions of collections etc.~at the beginning of this section
the run algorithm is inherently quantum, so the definition given here
differs from that in \cite{Backes:2004:Secure}. However, we try to
capture the structure and the behaviour of the scheduling.

Explanations on the individual steps of the algorithm below can be
found after the definition.

\begin{definition}[Run]\label{def:run}\index{run}
  Let a closed collection $\Hat C$ and some security parameter
  $k\in\setN$ be given. Let $\mathsf X$ denote the master scheduler of
  $\Hat C$.
  
  For any machine $\mathsf M\in[\Hat C]$ let
  $$
  \calH_\mathsf M:=
  \setC^{\mathit{QStates}_\mathsf M}\otimes
  \setC^{\mathit{CStates}_\mathsf M}\otimes
  \setC^{\calI_\mathsf M}\otimes
  \setC^{\calO_\mathsf M}.
  $$
  $\calH_\mathsf M$ it the space accessible to $\mathsf M$,
  including the space of its inputs and outputs.
  
  Note that $\calI_\mathsf M$ has the structure $\calI_\mathsf M=
  \prod_{\port p\in\mathsf{in}(\mathit{Ports}_M)}\Sigma^*$, therefore
  $\setC^{\calI_\mathsf M}$ decomposes into
  $$
  \setC^{\calI_\mathsf M} =
  \bigotimes_{\port p\in\mathsf{in}(\mathit{Ports}_M)}
  \calH_{\port p\relax}
  \qquad\text{with}\qquad
  \calH_{\port p\relax} := \setC^{\Sigma^*}
  $$
  Below we will sometimes refer to these subsystems directly via
  their name~$\calH_{\port p\relax}$.
  
  Let further 
  $$
  \calH_{\Hat C} := \bigotimes_{\mathsf M\in\Hat C}.
  $$
  
  In the following, when we say that some operation $X$ (a
  superoperator or a measurement) is applied to some subsystem $\calH$
  of $\calH_{\Hat C}=:\calH_a\otimes\calH\otimes\calH_b$, we formally
  mean that $1\otimes X\otimes 1$ is applied to
  $\Rho(\calH_a\otimes\calH\otimes\calH_b)$ (the set of density
  operators on $\calH_{\Hat C}$). Here $1$ denotes the identity.

  Consider the following measurement process on $\calH_{\Hat C}$
  (formally, on the set $\Rho(\calH_{\Hat C})$ of density operators
  over $\calH_{\Hat C}$).
  \let\LABEL\label
  \begin{enumerate}
  \item\LABEL{run.ini} Prepare the state
    $$
    \bigotimes_{\mathsf M\in\Hat C} \rho^{\textit{ini},k}_{\mathsf M}
    \in \Rho(\calH_{\Hat C})
    $$
    where
    \begin{align*}
    \rho^{\textit{ini},k}_{\mathsf M} :=
    &\butter{\varepsilon}
    \,\otimes\,\butter{\mathsf 1^k} \\
    &\otimes\,\butter{\varepsilon,\dots,\varepsilon}\\
    &\otimes\,\butter{\varepsilon,\dots,\varepsilon}
    \in \Rho(\calH_{\mathsf M})
    \end{align*}
    (This means we initialised all machines to initial quantum state
    $\butter{\varepsilon}$, classical initial state $\mathsf 1^k$, and
    empty in- and output-spaces.)
  \item\LABEL{run.actmaster} Initialise the variable $\mathsf M_{CS}$
    (the \emph{current scheduler}) to have the value~$\mathsf X$.
    Prepare $\butter{\mathsf 1}$ in
    $\calH_{\masterclk}$.\footnote{Formally, we apply the
      superoperator $\rho\mapsto\butter{\mathsf 1}$ to
      $\calH_{\masterclk}$.}
  \item\LABEL{run.loop}\LABEL{run.pre.state} Perform a complete von-Neumann measurement in the
    computational basis (called a complete measurement from now on) on
    $\setC^{\mathit{CStates}_{\mathsf M_{CS}}}$. Let $s$ denote the outcome.
  \item\LABEL{run.terminate} If $s\in\textit{Fin}_\MCS$ and $\MCS=\mathsf X$,
    exit (the run is complete). If $s\in\textit{Fin}_\MCS$, but
    $\MCS\neq \mathsf X$, proceed to Step~\ref{run.actmaster}.
  \item\LABEL{run.truncate} For each port $\mathsf p\in\mathsf{in}(\Ports_\MCS)$
    s.t.~$l_\MCS(s,\mathsf p)=0$, prepare $\butter{\varepsilon}$
    in~$\calH_\mathsf p$.
  \item\LABEL{run.classical.in}
    For each $\mathsf p\in\mathsf{in}(\CPorts_\MCS)$, perform a complete
    measurement on $\calH_{\mathsf p}$. Let the outcome be $I_{\mathsf
      p}$.
  \item\LABEL{run.noinput} For each port $\mathsf
    p\in\mathsf{in}(\Ports_\MCS)$, measure whether $\calH_\mathsf p$
    is in state $\ket{\varepsilon}$ (whether it is empty). If all
    ports were empty, proceed to Step~\ref{run.actmaster}. Otherwise
    let $P$ be the set of the ports that were nonempty.
  \item\LABEL{run.switch} Switch the current scheduler, i.e.~apply the
    state-transition operator~$\Delta_\MCS$ to $\calH_\MCS$.
  \item\LABEL{run.poststate.classical.out} Perform a complete
    von-Neumann measurement in the computational basis (called a
    complete measurement from now on) on
    $\setC^{\mathit{CStates}_\MCS}$. Let $s'$ denote the outcome.  For
    each $\mathsf p\in\mathit{out}(\CPorts_\MCS)$, perform a complete
    measurement on $\calH_{\mathsf p}$. Let the outcome be $O_{\mathsf
      p}$.
  \item\LABEL{run.addtrace} Let $I:=(I_\mathsf p)_{\mathsf
      p\in\mathit{in}(\CPorts_\MCS)}$ and $O:=(O_\mathsf p)_{\mathsf
      p\in\mathit{out}(\CPorts_\MCS)}$.  Add
    $(\textit{name}_\MCS,s,I,s',O,P)$ to the trace
    (which initially is empty).
  \item\LABEL{run.to.buffer} Let $\mathrm{MOVE}$ denote the
    superoperator over $\setC^{\Sigma^*}\mathop{\!\otimes}
    \setC^{\Sigma^*}=:\calH_1\otimes\calH_2$ defined by
    $\rho\mapsto\butter{\varepsilon}\otimes\tr_2\rho$.\footnote{I.e.,~the
      second subsystem is prepared to be \butter\varepsilon, and
      then then the subsystems are swapped.} Then for each simple
    out-port $\port p!\in\Ports_{\mathsf M_{CS}}$ perform the
    following: Measure, whether $\port p!$ is empty, i.e.~measure
    whether $\calH_{\port p!}$ is in state $\ket{\varepsilon}$. If
    nonempty, apply $\mathrm{MOVE}$ to $\calH_{\port
      p!}\otimes\calH_{\port p-?}$.  Then (if $\port p!$ was
    nonempty) switch buffer $\buff p$, i.e.~apply $\Deltabuff$ to
    $\calH_{\buff p}$.
  \item\LABEL{run.clear.ports} For each $\mathsf p\in\Ports_{\mathsf
      M_{CS}}$, prepare $\butter\varepsilon$ in $\calH_\mathsf p$.
    (With $\varepsilon$ being the empty word.)
  \item\LABEL{run.findsched} Let $\port s<!$ be the first clock
    out-port from $\CPorts_{\MCS}$ (in the ordering given by the port
    sequence $\Ports_{\MCS}$) with $I_{\port s<!}\neq\varepsilon$. If
    there is no such port, proceed to Step~\ref{run.actmaster}.
  \item\LABEL{run.schedbuff} Prepare $\butter{I_{\port s<!}}$ in
    $\calH_{\port s<?}$. Then switch buffer $\buff s$ (apply
    $\Deltabuff$ to $\calH_{\buff s}$).  Measure whether $\port s-!$
    is empty (measure whether $\calH_{\port s-!}$ is in state
    $\ket{\varepsilon}$).  If it is empty, proceed to
    Step~\ref{run.actmaster}.
  \item\LABEL{run.from.buffer} Apply $\mathrm{MOVE}$ to $\calH_{\port
      s-!}\otimes\calH_{\port s?}$.  Let $\MCS$ be the unique machine
    with $\port s?\in\Ports_{\MCS}$. Proceed to Step~\ref{run.loop}.
  \end{enumerate}
  
  This measurement process induces a probability distribution on the
  trace (see Step~\ref{run.addtrace}). We call this probability
  distribution $\run_{\Hat C,k}$ (the \emph{run/trace\index{trace} of $\Hat C$ on
    security parameter $k$}). We will also use $\run_{\Hat C,k}$ for a
  random variable with that distribution.
\end{definition}

We will now comment on the individual steps of the measurement process
above.

In Step~\ref{run.ini}, the initial states of the machines are
prepared. It consists of the security parameter $k$ (encoded as
$\mathsf 1^k$) as the classical state, and the empty word
$\ket\varepsilon$ as the quantum state.

Then, in Step~\ref{run.actmaster} the master scheduler is chosen to be
the next machine to be activated, and its $\masterclk$ port gets the
content $\mathsf 1$. The loop will jump to this step whenever the
activation token is lost, see below.

In Step~\ref{run.pre.state}, the classical state $s$ of the current
scheduler $\MCS$ (the machine to be activated in this iteration of the
loop) is measured for inclusion in the trace
(Step~\ref{run.addtrace}). So $s$ is the state \emph{before
  activation}.

In Step~\ref{run.terminate} it is checked, whether the current
scheduler is in a final state. If so, the master scheduler must be
activated, so we go back to Step~\ref{run.actmaster}. If the current
scheduler is the master scheduler and has terminated, the whole
process terminated and the run is complete.

In Step~\ref{run.truncate}, for each in-port the length function is
evaluated, and if it is $0$, the content of that port erased (so that
messages on this port are ignored).

Then, in Step~\ref{run.classical.in} the contents of all classical
ports are measured for inclusion in the trace
(Step~\ref{run.addtrace}).

In Step~\ref{run.noinput} it is checked, whether there is at least one
port containing data. If not, the current scheduler is not activated,
and the master scheduler is activated by going to
Step~\ref{run.actmaster}.

Step~\ref{run.switch} is probably the most important step in the run.
Here the current scheduler's state-transition operator is finally
applied.

Then in Step~\ref{run.poststate.classical.out} the classical state
$s'$ of the current scheduler and its classical outputs are measured
for inclusion into the trace.

In Step~\ref{run.addtrace} the classical state of the master scheduler
before and after execution, and its classical in- and outputs are
appended to a variable called the trace. This variable describes the
observable behaviour of the network and its final value (or the
possibly infinite sequence if the loop does not terminate) gives rise
to a probability distribution of observable behaviour, the \emph{run}
or \emph{trace on security parameter $k$}: $\run_{\Hat C,k}$.

Then, in Step~\ref{run.to.buffer}, for each simple out-port of $\MCS$,
that is nonempty, the content of this port is moved to the
corresponding buffer in-port. Then the buffer is activated, with the
effect that it stores the incoming message into its queue
(cf.~Definition~\ref{def:buffer}).

Now, in Step~\ref{run.clear.ports}, the contents of the ports of
$\MCS$ are erased (the contents of the clock out-ports are still
needed, but they have been measured above are therefore are still
accessible via the variables $O_{\port p<!}$.

In Step~\ref{run.findsched} we choose the first clock-out port of
$\MCS$ that contained output. If there is not such clock-out port, it
means that $\MCS$ did not want to schedule any connection, so the
master scheduler is activated again via Step~\ref{run.actmaster}.

Then in Step~\ref{run.schedbuff}, if some connection is to be
scheduled, the corresponding buffer is activated with the number of
the message as input on its clock in-port $\port p<?$. This has the
effect of moving the message from the queue to the buffer out-port.

Finally, in Step~\ref{run.from.buffer}, the message is moved from the
buffer's out-port to the recipients corresponding simple in-port, and
the recipient is noted to be the next machine to be activated (current
scheduler). Then the loop proceeds with Step~\ref{run.loop} (not
Step~\ref{run.actmaster}, since this would activate the master
scheduler).

We encourage the reader to compare this formal definition with
intuitive description of the scheduling given in
Section~\ref{sec:qns}.

So finally we got a random variable $\run_{\Hat C,k}$ describing the
observable behaviour of the network and can proceed to the next
section, where the actual security definitions will be stated.

%%% Local Variables:
%%% mode: latex
%%% coding: latin-1
%%% TeX-master: "quantum-security"
%%% End:
%%% Local IspellDict: british

\section{Quantum security definitions}
\label{sec:secdef}

In the preceding section we have defined what a quantum network is
(formally, as closed collection $\Hat M$). Further we have described
the evolution of such a network over time, and the defined a random
variable $\run_{\Hat M,k}$ representing the observable behaviour of
such a network when the security parameter is $k$. Using these
prerequisites, it is now easy to define quantum simulatable security.
In particular, nothing specially related to the quantum nature of our
protocols has to be taken into account any more, so most of this
section is very similar to \cite{Backes:2004:Secure}.

\subsection{Protocols}\index{protocol}\label{sec:protocols}

Remember, that we ``defined'' simulatable security in
Section~\ref{sec:whatis} approximately as follows (we will talk of an
ideal protocol instead of the special case of a trusted host here, for
greater generality):

A real protocol $\pi$ is as secure as an ideal protocol $\rho$ if
there for each real adversary $\sfAr$ there is a simulator $\sfAs$
s.t.~for all honest users $\sfH$ the view of $\sfH$ in runs with the
real protocol and real adversary is indistinguishable from runs with
the ideal protocol and the simulator.

The first point of this ``definition'' we will elaborate on, is the
notion of a protocol. From Section~\ref{sec:run} we already have the
notion of a collection. We remember that a non-closed collection is
one where some ports (the free ports) are still unconnected. So such
an open collection can be regarded as a protocol, where the in- and
outputs of the protocols go over the free ports to some still to be
specified outer world.

\begin{figure}[thb]
  \begin{center}
    \setlength{\unitlength}{4144sp}%
\begingroup\makeatletter\ifx\SetFigFont\undefined%
\gdef\SetFigFont#1#2#3#4#5{%
  \reset@font\fontsize{#1}{#2pt}%
  \fontfamily{#3}\fontseries{#4}\fontshape{#5}%
  \selectfont}%
\fi\endgroup%
\begin{picture}(5694,1104)(529,-1153)
\thinlines
{\color[rgb]{0,0,0}\put(991,-571){\framebox(300,150){}}
}%
{\color[rgb]{0,0,0}\put(1141,-571){\framebox(300,150){}}
}%
{\color[rgb]{0,0,0}\put(1066,-571){\framebox(300,150){}}
}%
{\color[rgb]{0,0,0}\put(1216,-571){\framebox(75,150){}}
}%
{\color[rgb]{0,0,0}\put(3151,-571){\framebox(300,150){}}
}%
{\color[rgb]{0,0,0}\put(3301,-571){\framebox(300,150){}}
}%
{\color[rgb]{0,0,0}\put(3226,-571){\framebox(300,150){}}
}%
{\color[rgb]{0,0,0}\put(3376,-571){\framebox(75,150){}}
}%
{\color[rgb]{0,0,0}\put(5401,-571){\framebox(300,150){}}
}%
{\color[rgb]{0,0,0}\put(5551,-571){\framebox(300,150){}}
}%
{\color[rgb]{0,0,0}\put(5476,-571){\framebox(300,150){}}
}%
{\color[rgb]{0,0,0}\put(5626,-571){\framebox(75,150){}}
}%
{\color[rgb]{0,0,0}\put(541,-511){\vector( 1, 0){450}}
}%
{\color[rgb]{0,0,0}\put(1441,-511){\vector( 1, 0){360}}
}%
{\color[rgb]{0,0,0}\put(3601,-511){\vector( 1, 0){450}}
}%
{\color[rgb]{0,0,0}\put(5851,-511){\vector( 1, 0){360}}
}%
{\color[rgb]{0,0,0}\multiput(1261,-1141)(0.00000,120.00000){5}{\line( 0, 1){ 60.000}}
\put(1261,-601){\vector( 0, 1){0}}
}%
{\color[rgb]{0,0,0}\multiput(3421,-1141)(0.00000,120.00000){5}{\line( 0, 1){ 60.000}}
\put(3421,-601){\vector( 0, 1){0}}
}%
{\color[rgb]{0,0,0}\multiput(5671,-1141)(0.00000,120.00000){5}{\line( 0, 1){ 60.000}}
\put(5671,-601){\vector( 0, 1){0}}
}%
{\color[rgb]{0,0,0}\put(1801,-961){\framebox(900,900){}}
}%
{\color[rgb]{0,0,0}\put(2701,-511){\vector( 1, 0){450}}
}%
{\color[rgb]{0,0,0}\put(4051,-961){\framebox(900,900){}}
}%
{\color[rgb]{0,0,0}\put(4951,-511){\vector( 1, 0){450}}
}%
\put(1171,-331){\makebox(0,0)[lb]{\smash{\SetFigFont{12}{14.4}{\rmdefault}{\mddefault}{\updefault}{\color[rgb]{0,0,0}$\buff{in}$}%
}}}
\put(3241,-331){\makebox(0,0)[lb]{\smash{\SetFigFont{12}{14.4}{\rmdefault}{\mddefault}{\updefault}{\color[rgb]{0,0,0}$\buff{net}$}%
}}}
\put(5491,-331){\makebox(0,0)[lb]{\smash{\SetFigFont{12}{14.4}{\rmdefault}{\mddefault}{\updefault}{\color[rgb]{0,0,0}$\buff{out}$}%
}}}
\put(811,-781){\makebox(0,0)[lb]{\smash{\SetFigFont{12}{14.4}{\rmdefault}{\mddefault}{\updefault}{\color[rgb]{0,0,0}$\port{in}-?$}%
}}}
\put(1351,-781){\makebox(0,0)[lb]{\smash{\SetFigFont{12}{14.4}{\rmdefault}{\mddefault}{\updefault}{\color[rgb]{0,0,0}$\port{in}-!$}%
}}}
\put(5761,-781){\makebox(0,0)[lb]{\smash{\SetFigFont{12}{14.4}{\rmdefault}{\mddefault}{\updefault}{\color[rgb]{0,0,0}$\port{out}-!$}%
}}}
\put(2386,-601){\makebox(0,0)[lb]{\smash{\SetFigFont{12}{14.4}{\rmdefault}{\mddefault}{\updefault}{\color[rgb]{0,0,0}$\port{net}!$}%
}}}
\put(1846,-601){\makebox(0,0)[lb]{\smash{\SetFigFont{12}{14.4}{\rmdefault}{\mddefault}{\updefault}{\color[rgb]{0,0,0}$\port{in}?$}%
}}}
\put(2161,-286){\makebox(0,0)[lb]{\smash{\SetFigFont{12}{14.4}{\rmdefault}{\mddefault}{\updefault}{\color[rgb]{0,0,0}$\mathsf A$}%
}}}
\put(4411,-286){\makebox(0,0)[lb]{\smash{\SetFigFont{12}{14.4}{\rmdefault}{\mddefault}{\updefault}{\color[rgb]{0,0,0}$\mathsf B$}%
}}}
\put(4096,-556){\makebox(0,0)[lb]{\smash{\SetFigFont{12}{14.4}{\rmdefault}{\mddefault}{\updefault}{\color[rgb]{0,0,0}$\port{net}?$}%
}}}
\put(4636,-556){\makebox(0,0)[lb]{\smash{\SetFigFont{12}{14.4}{\rmdefault}{\mddefault}{\updefault}{\color[rgb]{0,0,0}$\port{out}!$}%
}}}
\put(2881,-781){\makebox(0,0)[lb]{\smash{\SetFigFont{12}{14.4}{\rmdefault}{\mddefault}{\updefault}{\color[rgb]{0,0,0}$\port{net}-?$}%
}}}
\put(5131,-781){\makebox(0,0)[lb]{\smash{\SetFigFont{12}{14.4}{\rmdefault}{\mddefault}{\updefault}{\color[rgb]{0,0,0}$\port{out}-?$}%
}}}
\put(3511,-781){\makebox(0,0)[lb]{\smash{\SetFigFont{12}{14.4}{\rmdefault}{\mddefault}{\updefault}{\color[rgb]{0,0,0}$\port{net}-!$}%
}}}
\put(1171,-1006){\makebox(0,0)[lb]{\smash{\SetFigFont{12}{14.4}{\rmdefault}{\mddefault}{\updefault}{\color[rgb]{0,0,0}$\port{in}<?$}%
}}}
\put(3286,-1006){\makebox(0,0)[lb]{\smash{\SetFigFont{12}{14.4}{\rmdefault}{\mddefault}{\updefault}{\color[rgb]{0,0,0}$\port{net}<?$}%
}}}
\put(5491,-1006){\makebox(0,0)[lb]{\smash{\SetFigFont{12}{14.4}{\rmdefault}{\mddefault}{\updefault}{\color[rgb]{0,0,0}$\port{out}<?$}%
}}}
\end{picture}
  \end{center}
  \caption{A simple protocol}
  \label{fig:simple.proto}
\end{figure}
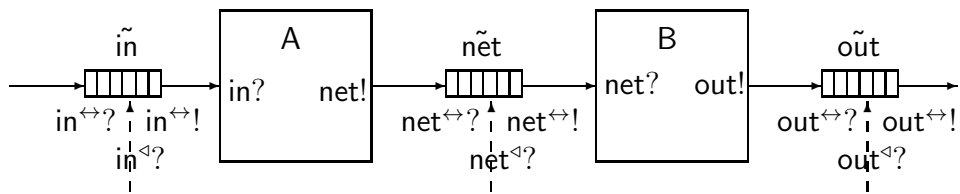

Let us consider an example: Two parties $\mathsf A$ and $\mathsf B$
form a protocol. They have a connection $\mathsf{net}$ between them
(so $\mathsf A$ and $\mathsf B$ have ports $\port{net}!$ and
$\port{net}?$, resp.), and for getting their in- and outputs they have
ports $\mathsf{in}_\mathsf A$ and $\mathsf{out}_\mathsf B$. Now the
collection describing that protocol would simply be $\Hat M:=\{\mathsf
A,\mathsf B\}$ (cf.~Figure~\ref{fig:simple.proto}). But when we look
at the free ports of $\Hat M$ (strictly spoken of the completion of
$\Hat M$), we note that
$$
\free([\Hat M]) = \{\port{in}-?,\port{out}-!,\port{in}<?,
\port{out}<?,\port{net}<?\}.
$$
This means, that some protocol user (honest user) would in
principle be able to connect e.g.~to the protocols
$\port{net}<?$-port. So a protocol user would ``see'' and even
``control'' the internal scheduling mechanisms of the protocol. But
this ability we would reserve for the adversary.

More precarious is the situation, if we would like to say that our
protocol implements some trusted host. Assume e.g.~that the trusted
host models some behaviour with an explicit insecurity (as is usual in
the modelling of trusted hosts, in most cases the adversary is at
least informed about the length of the data). Then the trusted host
(its completion) would have another free port called
e.g.~$\port{len}-!$. An honest user connecting to that port would of
course immediately ``notice a difference'', be it only that there are
more ports in the ideal than in the real protocol.

Therefore we need a way to specify which ports the honest user may
possibly connect to. Following \cite{Backes:2004:Secure} we will call
these ports the service ports, and a non-closed collection together
with the set of its service ports will be called a \emph{structure}. A
structure is our notion of a protocol. We cite the formal definition
of a structure from \cite{Backes:2004:Secure}:
\begin{definition}[Structures\index{structure} and service ports\biindex{service}{port} \cite{Backes:2004:Secure}]
  A \emph{structure} is a pair $(\Hat M,S)$ where $\Hat M$ is a
  collection of simple machines with $S\subseteq\free([\Hat M])$. The
  set $S$ is called \emph{service ports} of $(\Hat M,S)$.
\end{definition}

The notion of service ports allows us to specify the set of ports an
honest user must not connect to (we cite again):
\begin{definition}[Forbidden ports\biindex{forbidden}{port} \cite{Backes:2004:Secure}]
  For a structure $(\Hat M,S)$ let $\Bar S_{\Hat M}:=\free([\Hat
  M])\setminus S$. We call $\forb(\Hat M,S):=\Ports_{\Hat M}\cup\Bar
  S^c_{\Hat M}$ the \emph{forbidden ports}. (${}^c$ denotes the
  element-wise low-level complement.)
\end{definition}
It is sufficient to know, that $\forb(\Hat M,S)$ consists of the ports
$\Hat M$ may not have either because they would connect to non-service
ports, or because they are already used by the protocol and would give
rise to a name clash.

The next step is to specify, which honest users and adversaries are
valid ones. We must e.g.~disallow honest users which connect to
non-service ports. And we must guarantee that honest user, adversary
and structure together give rise to a closed collection, that can then
be executed (as in Section~\ref{sec:run}) to give rise to some
protocol trace. Here we can again cite \cite{Backes:2004:Secure}:

\begin{definition}[Configurations\index{configuration} \cite{Backes:2004:Secure}]
\begin{itemize}
\item A configuration of a structure $(\Hat M,S)$ is a tuple $\conf =
  (\Hat M,S,\sfH,\sfA)$ where
  \begin{itemize}
  \item $\sfH$ is a machine called \emph{user} (or \emph{honest user})\biindex{honest}{user}
    without forbidden ports, i.e.,~$\Ports_{\sfH}\cup\forb(\Hat
    M,S)=\emptyset$.
  \item $\sfA$ is a machine called \emph{adversary}.\index{adversary}
  \item the completion $\Hat C:=[\Hat M\cup\{\sfH,\sfA\}]$ is a closed
    collection.
  \end{itemize}
\item The set of configurations of $(\Hat M,S)$ is written $\Conf(\Hat
  M,S)$.
\item Let $(\Hat M_1,S)$ and $(\Hat M_2,S)$ be structures (with
  identical service ports). The set of \emph{suitable configurations}\biindex{suitable}{configuration}
  $\Conf^{\Hat M_2}(\Hat M,S)\subseteq\Conf(\Hat M_1,S)$ is defined by
  $(\Hat M_1,S,\sfH,\sfA)\in\Conf^{\Hat M_2}(\Hat M,S)$ iff
  $\Ports_{\sfH}\cap\forb(\Hat M_2,S)=\emptyset$.
\end{itemize}
\end{definition}

The first part of this definition tells us what honest users and
adversaries are admissible for some structure. Honest user $\sfH$ and
adversary $\sfA$ are admissible for the structure $(\Hat M,S)$ exactly
if $(\Hat M,S,\sfH,\sfA)\in\Conf(\Hat M,S)$.

The last part of the definition is needed further below. Consider an
honest user that is admissible for the real protocol $(\Hat M_1,S)$.
Then our security definition will use the same honest user also for
the ideal protocol $(\Hat M_2,S)$. If the honest user has ports that
would be forbidden with the ideal protocol, trouble is at hand.
Therefore the definition of suitable configurations additionally
requires that the honest user has no forbidden ports of either $(\Hat
M_1,S)$ nor $(\Hat M_2,S)$.

\subsection{The security relation}

Besides the unspecified notion of a ``protocol'', there was another
under-specified term in the ``definition'' from
Section~\ref{sec:whatis}: we required the view of $\mathsf H$ to be
indistinguishable in runs with real or ideal protocol. So let us first
specify what indistinguishability means in our scenario (we cite
again):

\begin{definition}[Small functions \cite{Backes:2004:Secure}]\biindex{small}{function}
  \begin{itemize}
  \item The class $\mathit{NEGL}$ of \emph{negligible functions}
    contains all functions $s:\setN\rightarrow\setR_{\geq0}$ that
    decrease faster than the inverse of every polynomial, i.e., for
    all positive polynomials $Q\ \exists k_0\ \forall k>k_0: s(k) <
    \tfrac{1}{Q(k)}$.
  \item The set $\mathit{SMALL}$ of functions
    $\setN\rightarrow\setR_{\geq0}$ is a \emph{class of small
      functions} if it is closed under addition, and with a function
    $g$ also contains every function $g':\setN\rightarrow
    \setR_{\geq0}$ with $g'\leq g$.
  \end{itemize}
\end{definition}

\begin{definition}[Indistinguishability \cite{Backes:2004:Secure}]\label{def:indist}\index{indistinguishable}
  Two families $(\mathsf{var})_{k\in\setN}$ and
  $(\mathsf{var}')_{k\in\setN}$ of probability distributions (or
  random variables) on common domains $(D_k)_{k\in\setN}$ are
  \begin{itemize}
  \item \emph{perfectly indistinguishable}\biindex{perfectly}{indistinguishable} (``='') iff $\forall
    k\in\setN:\mathsf{var}_k=\mathsf{var}'_k$.
  \item \emph{statistically indistinguishable}\biindex{statistically}{indistinguishable}
    (``$\approx_{\mathit{SMALL}}$'') for a class $\mathit{SMALL}$ of
    small functions if the distributions are discrete and their
    statistical distances, as a function of $k$, are small, i.e.
    $$
    \SD(\mathsf{var}_k,\mathsf{var}'_k)_{k\in\setN}
    :=
    \bigl(
    \tfrac12
    \sum_{d\in D_k}
    \lvert \Pr(\mathsf{var}_k=d) - \Pr(\mathsf{var}'_k=d)\rvert
    \in\textit{SMALL}.
    $$
\end{itemize}
\end{definition}

Mostly we will use $\textit{SMALL}:=\textit{NEGL}$.

The last term in our definition that still has to be defined is that
of a view. Intuitively, a view is everything (classical) a machine
experiences during the run. Since in the definition of the run, every
record in the run is tagged with the name of the corresponding
machine, it is now easy to define the view of a machine $\mathsf M$ by
removing all entries not tagged with the name of that machine from the
run. Formally:

\begin{definition}[Views]\index{view}
  Let a closed collection $\Hat C$ be given.  Then
  Definition~\ref{def:run} gives rise to a family of random variables
  $\run_{\Hat C,k}$. Let further $\mathsf M\in\Hat C$ be a simple
  machine or a master scheduler. Then the \emph{view $\view_{\Hat
      C,k}$ of $\mathsf M$ on security parameter $k$} is the
  subsequence of $\run_{\Hat C,k}$ resulting by taking only the
  elements $(n,s,I,s',O,p)\in\run_{\Hat C,k}$ satisfying
  $n=\textit{name}_{\mathsf M}$.
  
  If $\conf=(\Hat M,S,\sfH,\sfA)$ is a configurations, then we define
  $\view_{\conf,k}(\mathsf M):=\view_{\Hat
    M\cup\{\sfH,\sfA\},k}(\mathsf M)$.
\end{definition}
  
We restate our informal ``definition''.

A real protocol $\pi$ is as secure as an ideal protocol $\rho$ if
there for each real adversary $\sfAr$ there is a simulator $\sfAs$
s.t.~for all honest users $\sfH$ the view of $\sfH$ in runs with the
real protocol and real adversary is indistinguishable from runs with
the ideal protocol and the simulator.

Using the definitional tools developed above, we can capture this
definition formally (slightly modified in comparison to
\cite{Backes:2004:Secure} to include the notion of \emph{strict
  statistical security}):

\begin{definition}[Security for structures]\label{def:sec}\biindex{simulatable}{security}
  Let structures $(\Hat M_1,S)$ and $(\Hat M_2,S)$ with identical sets
  of service ports be given.
  \begin{itemize}
  \item $(\Hat M_1,S)\geq_\SEC^{\textit{perf}}(\Hat M_2,S)$, spoken $(\Hat
    M_1,S)$ is \emph{perfectly as secure as}\biindex{perfect}{security} $(\Hat M_1,S)$, iff for
    every configurations $\conf_1=(\Hat
    M_1,S,\sfH,\sfAr)\in\Conf^{\Hat M_2}(\Hat M_1,S)$ (the real
    configuration), there exists a configuration $\conf_2=(\Hat
    M_2,S,\sfH,\sfAs)\in\Conf(\Hat M_2,S)$ with the same $\sfH$ (the
    ideal configuration) s.t.
    $$
    (\view_{\conf_1,k}(\sfH))_k =
    (\view_{\conf_2,k}(\sfH))_k.
    $$
  \item $(\Hat M_1,S)\geq_\SEC^{\textit{SMALL}}(\Hat M_2,S)$, spoken
    $(\Hat M_1,S)$ is \emph{strictly statistically as secure as}\triindex{strict}{statistical}{security} $(\Hat
    M_1,S)$, for a class of small functions, iff for every
    configurations $\conf_1=(\Hat M_1,S,\sfH,\sfAr)\in\Conf^{\Hat
      M_2}(\Hat M_1,S)$ (the real configuration), there exists a
    configuration $\conf_2=(\Hat M_2,S,\sfH,\sfAs)\in\Conf(\Hat
    M_2,S)$ with the same $\sfH$ (the ideal configuration) s.t.
    $$
    (\view_{\conf_1,k}(\sfH))_k \approx_{\textit{SMALL}}
    (\view_{\conf_2,k}(\sfH))_k.
    $$
  \end{itemize}
  
  In both cases, we speak of \emph{universal simulatability} (or
  \emph{universal security}) if $\sfAs$ in $\conf_2$ does not depend
  on $\sfH$ (only on $\Hat M_1$, $S$, and $\sfAr$), and we use the
  notation $\geq_\SEC^{\textit{perf}}$ etc.~for this.
\end{definition}

\subsection{Corruption}\index{corruption}

So far, we have not modelled the possibility of corrupting a party.
However, the approach of \cite{Backes:2004:Secure} applies to our
setting with virtually no modifications, so we simple refer to
\cite{Backes:2004:Secure}. (A very short sketch of their approach is
also found in Section~\ref{sec:corruption}.)

%%% Local Variables:
%%% mode: latex
%%% coding: latin-1
%%% TeX-master: "quantum-security"
%%% End:
%%% Local IspellDict: british

\section{Composition}
\label{sec:comp}

In the present section we are going to show the simple composition
theorem. See Section~\ref{sec:comp.flavours} for an overview what
this composition theorem is for, and different flavours of composition
theorems exist.

\subsection{Combinations}\label{sec:combi}

Consider a network consisting of some machines $\mathsf
M_1,\dots,\mathsf M_n$. Imagine now taking two of these machines (say
$\mathsf M_1,\mathsf M_2$) and putting them into a
cardboard\index{cardboard} box (without disconnecting any of the
network cables). Then (if the machines are not equipped with some
special hardware for detecting cardboard boxes, e.g.~a light sensor)
that no machine will detect a difference, i.e.~the views of all
machines are unchanged. Even more, we can now consider the cardboard
box as a new, more complex machine $\Comb(\mathsf M_1,\mathsf M_2)$
(the combination of $\mathsf M_1$, $\mathsf M_2$). The view of
$\Comb(\mathsf M_1,\mathsf M_2)$ then contains the views of $\mathsf
M_1$ and $\mathsf M_2$, so we can still claim that the view of no
machine (not even $\mathsf M_1$ or $\mathsf M_2$) is changed by
removing $\mathsf M_1$ and $\mathsf M_2$ from the network and
replacing them by $\Comb(\mathsf M_1,\mathsf M_2)$.

This seemingly trivial observation is---when formally modelled---a
powerful tool for reasoning about networks which we will need below.

Before we define the combination, consider the following technical
definition:

\begin{definition}[Canonisation]\index{canonisation}
  Let a master scheduler or a simple machine $\mathsf M$ with
  state-transition operator $\Delta_M$ be given.

  Then
  $$
  \calH_\mathsf M:=
  \setC^{\mathit{QStates}_\mathsf M}\otimes
  \setC^{\mathit{CStates}_\mathsf M}\otimes
  \setC^{\calI_\mathsf M}\otimes
  \setC^{\calO_\mathsf M}
  $$
  is the state space of $\mathsf M$ (together with its inputs and
  outputs).

%   Let $P_0$ be the projection onto
%   $\ket{\varepsilon,\dots,\varepsilon}$ on $\setC^{\calI_\mathsf M}$,
%   i.e.~$P_0=\butter{\varepsilon,\dots,\varepsilon}$. Then
%   $P:=\{P_0,P_1\}$ with $P_1:=1-P_0$ is the von-Neumann measurement on
%   $\setC^{\calI_\mathsf M}$, yielding outcome $0$ if all inputs are
%   empty and $1$ otherwise.

  Then we define $\Tilde\Delta_\mathsf M$ to be the superoperator
  resulting from the following measurement process on $\calH_\mathsf M$:
  \begin{itemize}
  \item Measure the classical state $s$ of $\mathsf M$.
  \item For each in-port of $\mathsf M$, measure whether it is empty.
  \item For each in-port $\mathsf p$ s.t.~$l_\mathsf M(c,\mathsf
    p)=0$, prepare $\butter{\varepsilon}$ in~$\calH_\mathsf p$.
  \item If $s\notin\Fin_\mathsf M$, and at least one port was nonempty,
    apply $\Delta_M$.
  \end{itemize}

  Then the \emph{canonisation of $\mathsf M$} is the machine $$\mathsf
  {\Tilde M}:= (\textit{name}_\mathsf M,\textit{Ports}_\mathsf M,
  \textit{CPorts}_\mathsf M, \textit{QStates}_\mathsf
  M,\textit{CStates}_\mathsf M, \Tilde\Delta_\mathsf M,l_\mathsf
  M,\textit{Fin}_\mathsf M),$$
  i.e.~$\mathsf{\Tilde M}$ results from
  $\mathsf M$ by replacing its state-transition operator by
  $\Tilde\Delta_\mathsf M$.
\end{definition}

Now note that in the run algorithm (Definition~\ref{def:run}) the only
place where the state-transition operator $\Tilde\Delta_\mathsf M$ can
be applied is Step~\ref{run.switch}. But preceding that step,
measurements occurred that guarantee that at least one port is
nonempty, that $s$ is not a final state, and that all ports with
$l_\mathsf M(c,\mathsf p)=0$ are empty.  Therefore there is no
difference between using $\Delta_\mathsf M$ and $\Tilde\Delta_\mathsf
M$ in Step~\ref{run.switch}. From this insight, the following lemma
follows:

\begin{lemma}\label{lemma:canon}
  Let $\Hat C$ be some closed collection and $\mathsf M\in\Hat C$ a
  master scheduler or a simple machine. Let $\mathsf{\Tilde M}$ be the
  canonisation of $\mathsf M$, and $\Hat D:=\Hat C\setminus\{\mathsf
  M\}\cup\{\mathsf{\Tilde M}\}$ result from replacing $\mathsf M$ by
  $\mathsf {\Tilde M}$.
  Then $\Hat D$ is closed, and for all $k\in\setN$
  $$
  \run_{\Hat C,k} = \run_{\Hat D,k}.
  $$
\end{lemma}

Using the canonisation it is now easy to define the combination of two
machines:

\begin{definition}[Combination]\index{combination}
  Let two machines $\mathsf M_1,\mathsf M_2$ with disjoint sets of
  ports be given. Assume both them to be either master schedulers or simple
  machines or a combinations thereof.

  Denote by $\mathsf{\Tilde M}_1$ and $\mathsf{\Tilde M}_2$ the canonisations of
  $\mathsf M_1$ and $\mathsf M_2$.

  Then let
  \begin{align*}
    \textit{name}_\mathsf C &:= \textit{name}_{\mathsf
      {\Tilde M}_1}\textit{name}_{\mathsf{\Tilde M}_2} \\
    \Ports_\mathsf C &:= \Ports_{\mathsf{\Tilde M}_1} \cup \Ports_{\mathsf{\Tilde M}_2} \\
    \CPorts_\mathsf C &:= \CPorts_{\mathsf{\Tilde M}_1} \cup \CPorts_{\mathsf{\Tilde M}_2} \\
    \QStates_\mathsf C &:= \QStates_{\mathsf{\Tilde M}_1} \times \QStates_{\mathsf{\Tilde M}_2} \\
    \CStates_\mathsf C &:= \CStates_{\mathsf{\Tilde M}_1} \times
    \CStates_{\mathsf{\Tilde M}_2} \\
    \Delta_\mathsf C &:= \Delta_{\mathsf
      M_1}\otimes\Delta_{\mathsf{\Tilde M}_2}\\
    l_\mathsf C((c_1,c_2),\port p) &:=
    \begin{cases}
      l_{\mathsf{\Tilde M}_1}(c_1,\port p), & \text{if }\port
      p\in\Ports_{\mathsf{\Tilde M}_1} \\
      l_{\mathsf{\Tilde M}_2}(c_2,\port p), & \text{if }\port
      p\in\Ports_{\mathsf{\Tilde M}_2}
    \end{cases} \\
    \Fin_{\mathsf C} &:= \Fin_{\mathsf{\Tilde M}_1}\times\Fin_{\mathsf{\Tilde M}_2}
%    \begin{cases}
%      \Fin_{\mathsf{\Tilde M}_1}\times\CStates_{\mathsf{\Tilde M}_2}, &
%      \text{if $\mathsf{\Tilde M}_1$ master scheduler} \\
%      \CStates_{\mathsf{\Tilde M}_1}\times\Fin_{\mathsf{\Tilde M}_2}, &
%      \text{if $\mathsf{\Tilde M}_2$ master scheduler} \\
%      \Fin_{\mathsf{\Tilde M}_1}\times\Fin_{\mathsf{\Tilde M}_2}, &
%      \text{otherwise}
%    \end{cases}
  \end{align*}
  and
  $$
  \Comb(\mathsf M_1,\mathsf M_2) :=
  (\textit{name}_\mathsf C,\textit{Ports}_\mathsf C,\textit{CPorts}_\mathsf C,
  \textit{QStates}_\mathsf C,\textit{CStates}_\mathsf C,
  \Delta_\mathsf C,l_\mathsf C,\textit{Fin}_\mathsf C).
  $$
\end{definition}

We can now state the following combination lemma:
\begin{lemma}[Combination]\label{lemma:comb}%
  Let a closed collection $\Hat C$ be given. Assume $\mathsf M_1,
  \mathsf M_2\in\Hat C$ to be master schedulers, simple machines or a
  combination of both. Let $\Hat D:=\Hat C\setminus\{\mathsf
  M_1,\mathsf M_2\}\cup\{\Comb(\mathsf M_1,\mathsf M_2)\}$.

  Then for any machine $\mathsf M\in\Hat C\setminus\{\mathsf
  M_1,\mathsf M_2\}$ it is for all $k\in\setN$
  $$
  \view_{\Hat C}(\mathsf M) = \view_{\Hat D}(\mathsf M).
  $$

  Further let $\view_{\Hat D}(\mathsf M_1)$ denote the view of
  $\mathsf M_1$ extracted from the $v:=\view_{\Hat D}(\mathsf M_1)$ in
  the following manner:

  For each element $v_i=(\textit{name},(s_1,s_2),(I_1,I_2),(s_1',s_2'),(O_1,O_2),P)$,
  do the following
  \begin{itemize}
  \item If $P\cap\Ports_{\mathsf M_1}\neq\emptyset$, replace $v_i$ by
    $(\textit{name}_{\mathsf M_1},s_1,I_1,s_1',O_1',P)$.
  \item Otherwise, remove $v_i$ from the view.
  \end{itemize}

  Then
  $$
  \view_{\Hat C,k}(\mathsf M_1) = \view_{\Hat D,k}(\mathsf M_1).
  $$

  Then same holds for $\mathsf M_2$.
\end{lemma}

The proof of this lemma consists of first applying
Lemma~\ref{lemma:canon} to show $$
\run_{\Hat C,k}(\mathsf M) =
\run_{\Hat C',k}(\mathsf M).  $$
where $\Hat C'$ is $\Hat C$ with
$\mathsf M_1$ and $\mathsf M_2$ replaced by their canonisations. The
rest is a straightforward but long and tedious checking of each step
of the run-algorithm in Definition~\ref{def:run} to show $$
\view_{\Hat C}(\mathsf M) = \view_{\Hat D}(\mathsf M).  $$
A proof is given in Appendix~\ref{sec:proof.comb}.

So this lemma states that we can in fact replace any two machines by
their combination, without changing the behaviour of the network or
the views of the individual machines.

We write $\Comb(\mathsf M_1,\mathsf M_2,\dots,\mathsf M_n)$ for
$\Comb(\mathsf M_1,\Comb(\mathsf M_2,\Comb(\dots,\Comb(\mathsf
M_{n-1},\mathsf M_n)\dots)))$ to get a combination of more than two machines.

\subsection{Transitivity}\index{transitivity}

\begin{lemma}[Transitivity]
  Let $(\Hat M_1,S)\geq(\Hat M_2,S)\geq(\Hat M_3,S)$. Then $(\Hat
  M_1,S)\geq(\Hat M_3,S)$.

  Here $\geq$ may denote perfect, strict statistical, universal
  perfect and universal strict statistical security.
\end{lemma}

This lemma is obvious from Definition~\ref{def:sec}.

\subsection{The simple composition theorem}

In the preceding sections we have tried to get a strong notational and
structural similarity to \cite{Backes:2004:Secure}. We can now harvest
the fruits of this program: the definition of simple composition, the
simple composition theorem and the proof thereof are almost identical to
those in the RS framework. We restate the definition of composition
for self-containment:

\begin{definition}[Composition \cite{Backes:2002:Cryptographically}]
  Structures $(\Hat M_1,S),\dots,(\Hat M_n,S_n)$ are \emph{composable}
  if no port of $\Hat M_i$ is contained in $\forb(\Hat M_j,S_j)$ for
  $i\neq j$, and $S_1\cap\free([\Hat M_2])=S_2\cap\free([\Hat M_1])$.

  Their \emph{composition} is then $(\Hat M_1,S_1)\Vert\dots\Vert(\Hat
  M_n,S_n):=(\Hat M,s)$ with $\Hat M=\Hat M_1\cup\dots\cup\Hat M_n$
  and $S=(S_1\cup\dots\cup S_n)\cap\free([\Hat M])$.
\end{definition}

For details on the conditions in the definition of composability, see
\cite{Backes:2002:Cryptographically}.

We can now state the simple composition theorem (called \emph{Secure
  Two-system Composition} in \cite{Backes:2002:Cryptographically})
\begin{theorem}[Simple composition]\biindex{simple}{composition}\label{theo:comp}
  Let $(\Hat M_0,S_0)$, $(\Hat M_0',S_0)$ and $(\Hat M_1,S_1)$ be
  structures, s.t.~$(\Hat M_0,S_0)$, $(\Hat M_1,S_1)$ are composable,
  and $(\Hat M_0',S_0)$ and $(\Hat M_0',S_0)$ are composable.  Assume
  further $\textsf{ports}(\Hat M_0')\cap S_1^c=\textsf{ports}(\Hat
  M_0)\cap S_1^c$. ($\textsf{ports}(\Hat M)$ is the set of all ports
  of all machines in $\Hat M$)

  Then
  $$
  (\Hat M_0,S_0)\geq(\Hat M_0',S_0)
  \qquad
  \Longrightarrow
  \qquad
  (\Hat M_0,S_0)\Vert(\Hat M_1,S_1)\geq(\Hat M_0',S_0)\Vert(\Hat M_1,S_1)
  $$
  Here $\geq$ may denote perfect, strict statistical, universal
  perfect and universal strict statistical security.
\end{theorem}

The proof of the composition theorem in \cite{Pfitzmann:2001:Model} is
completely based on a higher-level view on the network model in the
sense that every statement about the view of machines is derived
through the combination lemma. So the proof of
\cite{Pfitzmann:2001:Model} applies in the quantum setting. (In fact
the proof in \cite{Pfitzmann:2001:Model} covers the more general case
of the security of systems, however, the composition theorem stated
above is just a special case of the composition theorem in
\cite{Pfitzmann:2001:Model}.) We therefore refer the reader to the
proof in \cite{Pfitzmann:2001:Model}.

%%% Local Variables:
%%% mode: latex
%%% coding: latin-1
%%% TeX-master: "quantum-security"
%%% End:
%%% Local IspellDict: british

% LocalWords:  superoperator

\section{Conclusions}
\label{sec:concl}\index{conclusions}

In the present work we have seen how to ``lift'' a classical model to
a quantum one. However, much possible work still lies ahead:

\begin{itemize}
\item\textit{Simplicity.}\index{simplicity} In the personal opinion of the author, the
  most urgent matter is the search for a model that is both simple
  (meaning both being simple to understand, and simple to use in
  proofs) and general (so taking recourse to restricting the
  possibilities of scheduling and message delivery would not be a
  solution).
  
  We believe that the complexity and the amount of details in the
  present work (the reader probably noticed them) is mostly due to the
  use of a message driven scheduling. A glance on the run-algorithm
  (Definition~\ref{def:run}) shows, that most of the steps are
  actually concerned with finding out which machine is to be activated
  with which inputs. Only one Step~\ref{run.switch} actually executes
  a machine's program.

  Note however that these complications are not particular to our
  model. Both in \cite{Backes:2004:Secure} and
  \cite{Canetti:2001:Security} most of the modelling is concerned with
  the order of activation of the machines. An effect of this is that
  security proofs tend to either get complicated and unreadable, or
  tend to ignore the details of scheduling almost completely and
  assume that message delivery will take place in a well-behaved and
  intuitive way.

  In the quantum case this problem is amplified by the fact that here
  one has to take care to explicitly specify any measurements done,
  instead of just referring to facts about the state of the system as
  in the classical case.
\item\textit{Formal proofs and machine
    verification.}\biindex{formal}{proof}\biindex{machine}{verification} Even when the security
  models have reached a point where proofs may concentrate on the
  essentials, large protocols may still be quite complex to manage. It
  would therefore be very helpful to have a hand tools for formally
  proving (using e.g.~rewriting rules for networks or similar)
  security, and for verifying (or even generating) proof with a
  machine.

  Some effort have already been done in that direction in the RS
  framework, e.g.~already in~\cite{Backes:2002:Deriving} the security
  of a protocol was shown in the theorem prover PVS.
\item\textit{Concrete security proofs.}\biindex{security}{proof} So
  far only a few protocols have been shown to be secure in a model of
  simulatable security.  E.g.,~the only family of quantum protocols to
  far is that of quantum key distribution.
\end{itemize}

\subsection{Acknowledgements}

We thank Dennis Hofheinz and Jörn Müller-Quade for many productive
discussions on various topics in the field of simulatable security.
This work was founded by the European Research Project
\textsc{ProSecCo} IST-2001-39227.

%%% Local Variables:
%%% mode: latex
%%% coding: latin-1
%%% TeX-master: "quantum-security"
%%% End:
%%% Local IspellDict: british

\begin{appendix}
  
\section{Postponed proofs}
\label{sec:proofs}

\subsection{Combination lemma (Lemma~\ref{lemma:comb})}
\biindex{proof of}{combination lemma}\label{sec:proof.comb}

\def\TM#1{{\Tilde{\mathsf M}_{#1}}}
\def\MCSX{{\Tilde{\mathsf M}_{CS}}}
\def\COMB{{\Comb(\mathsf M_1,\mathsf M_2)}}

Let $\Tilde C:=\Hat C\setminus\{\mathsf M_1,\mathsf
M_2\}\cup\{\mathsf{\Tilde M}_1,\mathsf{\Tilde M}_2\}$, i.e.~$\Tilde C$
results from $\Hat C$ by replacing $\mathsf M_i$ by their
canonisations $\mathsf{\Tilde M}_i$. Then by Lemma~\ref{lemma:canon},
we have
$$
\view_{\Hat C}(\mathsf M) = \view_{\Tilde C}(\mathsf M).
$$
So for proving the first part of Lemma~\ref{lemma:comb} it is
sufficient to show
\begin{equation}\label{eq:proof.comb.canon}
\view_{\Tilde C}(\mathsf M) = \view_{\Hat D}(\mathsf M).
\end{equation}

In order to do this, we will start with the run-algorithm for $\Tilde
C$ and perform a series of rewritings, until we have reached the
run-algorithm for $\Hat D$.

Let $A_0$ denote the run-algorithm for $\Tilde C$. Then we derive the
algorithm $A_1$ by replacing Steps~\ref{run.actmaster}
and~\ref{run.from.buffer} by
\begin{enumerate}\itshape
\item[\ref{run.actmaster}'.] Initialise the variable $\mathsf M_{CS}$
  (the \emph{current scheduler}) to have the value~$\mathsf X$.
  Prepare $\butter{\mathsf 1}$ in $\calH_{\masterclk}$.  Set
  $\MCSX:=\Tilde{\mathsf X}$. (Here $\mathsf{\Tilde X}:=\COMB$ if
  $\TM1$ or $\TM2$ is master scheduler, and $\mathsf{\Tilde
    X}:=\mathsf X$ otherwise.)
\item[\ref{run.from.buffer}'.] Apply $\mathrm{MOVE}$ to $\calH_{\port
    s-!}\otimes\calH_{\port s?}$.  Let $\MCS$ be the unique machine in
  $\Tilde C$ with $\port s?\in\Ports_{\MCS}$. Let $\MCSX$ be the
  unique machine in $\Hat D$ with $\port s?\in\Ports_{\MCS}$.
  Proceed to
  Step~\ref{run.loop}.

  If $\MCS\in\{\TM1,\TM2\}$, assign $\MCSX:=\COMB$, otherwise set
  $\MCSX:=\MCS$.
\end{enumerate}

In both steps we have only set a hitherto unused variable, but changed
nothing else, so the behaviour of $A_0$ and $A_1$ are identical,
i.e.~if we define $O_i$ to be the output (the trace) of $A_i$,
\begin{equation}\label{eq:pr.comb.first}
\run_{\Tilde C}=O_0=O_1.
\end{equation}

Further note, that it now holds after every step of the algorithm, 
that $\MCSX=\MCS$ for $\MCS\notin\{\TM1,\TM2\}$, and $\MCSX=\COMB$
otherwise.

We now derive the algorithm $A_2$ from $A_1$ by replacing
Step~\ref{run.pre.state} by
\begin{enumerate}\itshape
\item[\ref{run.pre.state}'.] Perform a complete von-Neumann
  measurement in the computational basis (called a complete
  measurement from now on) on $\setC^{\mathit{CStates}_{\MCSX}}$. Let
  $\tilde s$ denote the outcome.
  
  If $\MCS\notin\{\TM1,\TM2\}$, set $s:=\tilde s$, otherwise let $s$
  be the $i$-th component of $\tilde s$ (for
  $\MCS=\TM i$).
\end{enumerate}

We use here 
$\setC^{\CStates_{\COMB}}:=\setC^{\CStates_{\TM1}}\otimes\setC^{\CStates_{\TM2}}$.

If $\MCS\notin\{\TM1,\TM2\}$, this step evidently behaves like
Step~\ref{run.pre.state} (except that it sets an variable that is not
used in other steps). If $\MCS=\TM1$, the classical state of $\TM2$ is
additionally measured, however, this state was already measured in
Step~\ref{run.poststate.classical.out}, so measuring it does not
disturb the state of the system.  Further the result the original
Step~\ref{run.pre.state} would have yielded ($s$) is reconstructed
from ($\tilde s$).  Therefore in this case Step~\ref{run.pre.state}'
behaves like Step~\ref{run.pre.state}.  Analogously for $\TM2$.  So we
have
\begin{equation}
O_1=O_2.
\end{equation}

Similarly, we get algorithm $A_3$ by successively replacing
Steps~\ref{run.truncate}, \ref{run.classical.in}, \ref{run.noinput},
\ref{run.poststate.classical.out}, \ref{run.to.buffer}, \ref{run.clear.ports},
\ref{run.findsched} by
\begin{enumerate}\itshape
\item[\ref{run.truncate}'.] For each port $\mathsf
  p\in\mathsf{in}(\Ports_\MCSX)$ s.t.~$l_\MCSX(\tilde s,\mathsf p)=0$,
  prepare $\butter{\varepsilon}$ in~$\calH_\mathsf p$.
  \item[\ref{run.classical.in}'.]
    For each $\mathsf p\in\mathsf{in}(\CPorts_\MCSX)$, perform a complete
    measurement on $\calH_{\mathsf p}$. Let the outcome be $I_{\mathsf
      p}$.
  \item[\ref{run.noinput}'.] For each port $\mathsf
    p\in\mathsf{in}(\Ports_\MCSX)$, measure whether $\calH_\mathsf p$
    is in state $\ket{\varepsilon}$ (whether it is empty). If all
    ports were empty, proceed to Step~\ref{run.actmaster}'. Otherwise
    let $P$ be the set of the ports that were nonempty.
  \item[\ref{run.poststate.classical.out}'.] Perform a complete
    von-Neumann measurement in the computational basis (called a
    complete measurement from now on) on
    $\setC^{\mathit{CStates}_\MCSX}$. Let $\tilde s'$ denote the outcome.  For
    each $\mathsf p\in\mathit{out}(\CPorts_\MCSX)$, perform a complete
    measurement on $\calH_{\mathsf p}$. Let the outcome be $O_{\mathsf
      p}$.

  If $\MCS\notin\{\TM1,\TM2\}$, set $s':=\tilde s'$, otherwise let $s'$
  be the $i$-th component of $\tilde s'$ (for
  $\MCS=\TM i$).
  \item[\ref{run.to.buffer}'.] For each simple
    out-port $\port p!\in\Ports_{\MCSX}$ perform the
    following: Measure, whether $\port p!$ is empty, i.e.~measure
    whether $\calH_{\port p!}$ is in state $\ket{\varepsilon}$. If
    nonempty, apply $\mathrm{MOVE}$ to $\calH_{\port
      p!}\otimes\calH_{\port p-?}$.  Then (if $\port p!$ was
    nonempty) switch buffer $\buff p$, i.e.~apply $\Deltabuff$ to
    $\calH_{\buff p}$.
  \item[\ref{run.clear.ports}'.] For each $\mathsf
    p\in\Ports_{\MCSX}$, prepare $\butter\varepsilon$ in
    $\calH_\mathsf p$.
  \item[\ref{run.findsched}'.] Let $\port s<!$ be the first clock
    out-port from $\CPorts_{\MCSX}$ (in the ordering given by the port
    sequence $\Ports_{\MCSX}$) with $I_{\port s<!}\neq\varepsilon$. If
    there is no such port, proceed to Step~\ref{run.actmaster}'.
\end{enumerate}
Note: In all steps we have replaced $\MCS$ by $\MCSX$, in
Step~\ref{run.truncate} also $s$ by $\tilde s$, in
Step~\ref{run.poststate.classical.out} we have additionally generated
$s'$ from the measurement result $\tilde s'$.

The reader can easily convince himself (similarly to the reasoning on
the replacement of Step~\ref{run.classical.in} above), that each of
the replacements does not modify the behaviour of the algorithm, so
\begin{equation}
O_2=O_3.
\end{equation}

Now we replace Step~\ref{run.terminate} by the following, resulting in
an algorithm $A_4$
\begin{enumerate}\itshape
\item[\ref{run.terminate}'.] If $s\in\textit{Fin}_\MCS$ and
  $\MCS=\mathsf X$, exit (the run is complete). If
  $s\in\textit{Fin}_\MCS$, but $\MCS\neq \mathsf X$, erase
  $\calH_\port p$ for all $\port p\in\textsf{in}(\Ports_\MCSX)$, and
  proceed to Step~\ref{run.actmaster}.
\end{enumerate}
This does not change the behaviour of the algorithm, since if a
machine has terminated, its in-ports are not read any more, so they
can safely be erased. The in-port of all other machines are either
empty or will not be read any more, so they can also safely be erased.
Therefore
\begin{equation}
O_3=O_4.
\end{equation}

Now we replace Step~\ref{run.terminate} by the following, resulting in
an algorithm $A_5$
\begin{enumerate}\itshape
\item[\ref{run.terminate}''.] If $\tilde s\in\textit{Fin}_\MCSX$ and
  $\MCSX=\mathsf{\Tilde X}$, exit (the run is complete). If $\tilde
  s\in\textit{Fin}_\MCSX$, but $\MCSX\neq\mathsf{\Tilde X}$, erase
  $\calH_\port p$ for all $\port p\in\textsf{in}(\Ports_\MCSX)$, and
  proceed to Step~\ref{run.actmaster}'.
\end{enumerate}
Here $\mathsf{\Tilde X}:=\COMB$ if $\TM1$ or $\TM2$ is master
scheduler, and $\mathsf{\Tilde X}:=\mathsf X$ otherwise.

To comparing Steps~\ref{run.terminate}' and \ref{run.terminate}'' 
we have to distinguish the following cases:
\begin{itemize}
\item If $\MCS\notin\{\TM1,\TM2\}$, it is $\MCS=\MCSX$, so
  Steps~\ref{run.terminate}' and~\ref{run.terminate}'' exhibit the same
  behaviour.
\item If $\MCS=\TM1$ and $s\notin\Fin_{\TM1}$, it also is $\tilde
  s\notin\Fin_{\COMB}$, so 
  Steps~\ref{run.terminate}' and~\ref{run.terminate}'' exhibit the same
  behaviour.
\item If $\MCS=\TM1$, $s\in\Fin_{\TM1}$ and $\tilde s\in\Fin_{\COMB}$,
  Steps~\ref{run.terminate}' and~\ref{run.terminate}'' exhibit the same
  behaviour.
\item If $\MCS=\TM1$, $\TM1$ is not master scheduler,
  $s\in\Fin_{\TM1}$, and $\tilde s\notin\Fin_{\COMB}$, then
  Step~\ref{run.terminate}' will jump to Step~\ref{run.actmaster}',
  while from Step~\ref{run.terminate}'' the algorithm $A_5$ will
  proceed through Steps~\ref{run.truncate}', \ref{run.classical.in}' and
  \ref{run.noinput}'. Since $l(s,\port p)=0$ for all in-ports of $\TM1$
  (since $\TM1$ was canonised), in Step~\ref{run.truncate}' all inputs
  will be erased. So in Step~\ref{run.noinput}', algorithm $A_5$ will
  jump to Step~\ref{run.actmaster}'. So in this case, 
  Steps~\ref{run.terminate}' and~\ref{run.terminate}'' exhibit the same
  behaviour.
\item If $\MCS=\TM1$, $\TM1$ is master scheduler, $s\in\Fin_{\TM1}$,
  and $\tilde s\notin\Fin_{\COMB}$, then Step~\ref{run.terminate}'
  will cause the algorithm $A_4$ to terminate, while algorithm $A_5$
  will be caught in an infinite loop without output
  (Steps~\ref{run.actmaster}'--\ref{run.noinput}'), since
  $l_{\MCS}(s,\masterclk)=0$ (because $\TM1$ was normalised). So also in
  this case, the trace will end here.
\item For $\MCS=\TM2$, we distinguish analogous cases.
\end{itemize}
So we can conclude
\begin{equation}
  O_4=O_5.
\end{equation}

Further (using the same argument as above, where we replaced
Step~\ref{run.terminate} by \ref{run.terminate}'), we can replace
Step~\ref{run.terminate}'' by Step~\ref{run.terminate}''' and get
algorithm $A_6$:
\begin{enumerate}\itshape
\item[\ref{run.terminate}'''.] If $\tilde s\in\textit{Fin}_\MCSX$ and
  $\MCSX=\mathsf{\Tilde X}$, exit (the run is complete). If $\tilde
  s\in\textit{Fin}_\MCSX$, but $\MCSX\neq\mathsf{\Tilde X}$, 
  proceed to Step~\ref{run.actmaster}'.
\end{enumerate}
so
\begin{equation}
O_5=O_6.
\end{equation}

Now, we make the following change to Step~\ref{run.switch}, resulting
in algorithm $A_7$:
\begin{enumerate}\itshape
\item[\ref{run.switch}'.]  Apply the state-transition
  operator~$\Delta_\MCS$ to $\calH_\MCS$. If $\MCS=\TM1$, then
  additionally apply $\Delta_\TM2$ to $\calH_\TM2$.
  If $\MCS=\TM2$, then additionally apply
  $\Delta_\TM1$ to $\calH_\TM1$.
\end{enumerate}
If $\MCS\neq\TM1$, then either all in-ports of $\TM1$ are empty or
$\TM1$ is in a final state. Since $\TM1$ has been canonised, applying
$\Delta_\TM1$ in this case behaves like the identity. The same holds
for $\TM2$. So the above changes have no effect, and
\begin{equation}
  O_6=O_7.
\end{equation}
Now we replace Step~\ref{run.switch}' by~\ref{run.switch}'', yielding
algorithm $A_8$:
\begin{enumerate}\itshape
\item[\ref{run.switch}''.]  Apply the state-transition
  operator~$\Delta_\MCSX$ to $\calH_\MCSX$. 
\end{enumerate}
By definition of $\Delta_\MCSX$, and using
$\calH_{\COMB}:=\calH_{\TM1}\otimes\calH_{\TM2}$, we have
\begin{equation}
  O_7=O_8.
\end{equation}

Now we replace Step~\ref{run.addtrace} by \ref{run.addtrace}',
yielding algorithm $A_9$:
\begin{enumerate}\itshape
\item[\ref{run.addtrace}'.]  Let $I:=(I_\mathsf p)_{\mathsf
    p\in\mathit{in}(\CPorts_\MCSX)}$ and $O:=(O_\mathsf p)_{\mathsf
    p\in\mathit{out}(\CPorts_\MCSX)}$.  Add
  $(\textit{name}_\MCSX,\tilde s,I,\tilde s',O,P)$ to the trace (which
  initially is empty).
\end{enumerate}
Let further $f$ denote the function that element-wise applies the
following mapping to the run
$$
(\textit{name},s,I,s',O,P) \quad\longmapsto\quad
\begin{cases}
  (\textit{name},s,I,s',O,P), &
  \text{if }\textit{name}\neq\textit{name}_{\Comb(\mathsf M_1,\mathsf M_2)},\\
  (\textit{name}_{\Tilde{\mathsf M}_1},s_1,I_1,s_1',O_1,P), &
  \text{if }\textit{name}=\textit{name}_{\Comb(\mathsf M_1,\mathsf M_2)}\\
  &
  \text{and }P\cap\Ports_{\mathsf{\Tilde M}_1}\neq\emptyset, \\
  (\textit{name}_{\Tilde{\mathsf M}_2},s_2,I_2,s'_2,O_2,P), &
  \text{if }\textit{name}=\textit{name}_{\Comb(\mathsf M_1,\mathsf M_2)}\\
  &
  \text{and }P\cap\Ports_{\mathsf{\Tilde M}_2}\neq\emptyset, \\
\end{cases}
$$
Here $s_i$ and $s_i'$ denote the $i$-th component of $s$ and $s'$, resp.

It is now easy to verify that
\begin{equation}
O_8=f(O_9).
\end{equation}
(Note that $P\cap\Ports_\mathsf M\neq\emptyset$ only if
$\MCS=\mathsf M$.)

Since $\calH_{\COMB}=\calH_{\TM1}\otimes\calH_{\TM2}$, we can replace
$\calH_{\Tilde C}$ by the isomorphic space $\calH_{\Hat D}$, and
rewrite Step~\ref{run.ini} as follows, giving algorithm $A_{10}$:
\begin{enumerate}\itshape
\item[\ref{run.ini}'.] Prepare the state
    $$
    \bigotimes_{\mathsf M\in\Hat D} \rho^{\textit{ini},k}_{\mathsf M}
    \in \Rho(\calH_{\Hat D})
    $$
    where
    \begin{align*}
    \rho^{\textit{ini},k}_{\mathsf M} :=
    &\butter{\varepsilon}
    \,\otimes\,\butter{\mathsf 1^k} \\
    &\otimes\,\butter{\varepsilon,\dots,\varepsilon}\\
    &\otimes\,\butter{\varepsilon,\dots,\varepsilon}
    \in \Rho(\calH_{\mathsf M}).
    \end{align*}
\end{enumerate}
Then obviously
\begin{equation}
O_9=O_{10}.
\end{equation}

From $A_{10}$ we can now derive $A_{11}$ by removing all intructions
that set $\MCSX$, $\tilde s$ or $\tilde s'$. $A_{11}$ the has the
following description:
\begin{enumerate}\itshape
\item[\ref{run.ini}'.] Prepare the state
    $$
    \bigotimes_{\mathsf M\in\Hat D} \rho^{\textit{ini},k}_{\mathsf M}
    \in \Rho(\calH_{\Hat D})
    $$
    where
    \begin{align*}
    \rho^{\textit{ini},k}_{\mathsf M} :=
    &\butter{\varepsilon}
    \,\otimes\,\butter{\mathsf 1^k} \\
    &\otimes\,\butter{\varepsilon,\dots,\varepsilon}\\
    &\otimes\,\butter{\varepsilon,\dots,\varepsilon}
    \in \Rho(\calH_{\mathsf M}).
    \end{align*}
\item[\ref{run.actmaster}'.] 
  Prepare $\butter{\mathsf 1}$ in $\calH_{\masterclk}$.  Set
  $\MCSX:=\Tilde{\mathsf X}$.
\item[\ref{run.pre.state}'.] Perform a complete von-Neumann
  measurement in the computational basis (called a complete
  measurement from now on) on $\setC^{\mathit{CStates}_{\MCSX}}$. Let
  $\tilde s$ denote the outcome.
\item[\ref{run.terminate}'''.] If $\tilde s\in\textit{Fin}_\MCSX$ and
  $\MCSX=\mathsf{\Tilde X}$, exit (the run is complete). If $\tilde
  s\in\textit{Fin}_\MCSX$, but $\MCSX\neq\mathsf{\Tilde X}$, 
  proceed to Step~\ref{run.actmaster}'.
\item[\ref{run.truncate}'.] For each port $\mathsf
  p\in\mathsf{in}(\Ports_\MCSX)$ s.t.~$l_\MCSX(\tilde s,\mathsf p)=0$,
  prepare $\butter{\varepsilon}$ in~$\calH_\mathsf p$.
\item[\ref{run.classical.in}'.]
  For each $\mathsf p\in\mathsf{in}(\CPorts_\MCSX)$, perform a complete
  measurement on $\calH_{\mathsf p}$. Let the outcome be $I_{\mathsf
    p}$.
\item[\ref{run.noinput}'.] For each port $\mathsf
  p\in\mathsf{in}(\Ports_\MCSX)$, measure whether $\calH_\mathsf p$
    is in state $\ket{\varepsilon}$ (whether it is empty). If all
    ports were empty, proceed to Step~\ref{run.actmaster}'. Otherwise
    let $P$ be the set of the ports that were nonempty.
\item[\ref{run.switch}''.]  Apply the state-transition
  operator~$\Delta_\MCSX$ to $\calH_\MCSX$. 
  \item[\ref{run.poststate.classical.out}'.] Perform a complete
    von-Neumann measurement in the computational basis (called a
    complete measurement from now on) on
    $\setC^{\mathit{CStates}_\MCSX}$. Let $\tilde s'$ denote the outcome.  For
    each $\mathsf p\in\mathit{out}(\CPorts_\MCSX)$, perform a complete
    measurement on $\calH_{\mathsf p}$. Let the outcome be $O_{\mathsf
      p}$.
\item[\ref{run.addtrace}'.]  Let $I:=(I_\mathsf p)_{\mathsf
    p\in\mathit{in}(\CPorts_\MCSX)}$ and $O:=(O_\mathsf p)_{\mathsf
    p\in\mathit{out}(\CPorts_\MCSX)}$.  Add
  $(\textit{name}_\MCSX,\tilde s,I,\tilde s',O,P)$ to the trace (which
  initially is empty).
  \item[\ref{run.to.buffer}'.] For each simple
    out-port $\port p!\in\Ports_{\MCSX}$ perform the
    following: Measure, whether $\port p!$ is empty, i.e.~measure
    whether $\calH_{\port p!}$ is in state $\ket{\varepsilon}$. If
    nonempty, apply $\mathrm{MOVE}$ to $\calH_{\port
      p!}\otimes\calH_{\port p-?}$.  Then (if $\port p!$ was
    nonempty) switch buffer $\buff p$, i.e.~apply $\Deltabuff$ to
    $\calH_{\buff p}$.
  \item[\ref{run.clear.ports}'.] For each $\mathsf
    p\in\Ports_{\MCSX}$, prepare $\butter\varepsilon$ in
    $\calH_\mathsf p$.
  \item[\ref{run.findsched}'.] Let $\port s<!$ be the first clock
    out-port from $\CPorts_{\MCSX}$ (in the ordering given by the port
    sequence $\Ports_{\MCSX}$) with $I_{\port s<!}\neq\varepsilon$. If
    there is no such port, proceed to Step~\ref{run.actmaster}'.
  \item[\ref{run.from.buffer}'.] Apply $\mathrm{MOVE}$ to
    $\calH_{\port s-!}\otimes\calH_{\port s?}$.  Let $\MCSX$ be the
    unique machine in $\Hat D$ with $\port s?\in\Ports_{\MCS}$.
    Proceed to Step~\ref{run.loop}.
\end{enumerate}
It is
\begin{equation}
O_{10}=O_{11}.
\end{equation}

This algorithm $A_{11}$ is in fact the algorithm for the run of $\Hat
D$, so
\begin{equation}\label{eq:pr.comb.last}
O_{11}=\run_{\Hat D}.
\end{equation}

By equations (\ref{eq:pr.comb.first}--\ref{eq:pr.comb.last}) we have
$$
\run_{\Tilde C}=f(\run_{\Hat D}).
$$

From this it follows for all $\mathsf M\in\Tilde C$,
$$
\view_{\Tilde C}(\mathsf M)=\view_{\Tilde C}(\mathsf M),
$$
with $\view_{\Tilde C}(\TM i)$ ($i=1,2$) defined as in Lemma~\ref{lemma:comb}.

Using \eqref{eq:proof.comb.canon}, the statement of the lemma follows.\qed

%\subsection{Simple composition theorem (Theorem~\ref{theo:comp})}
%\triindex{proof of}{simple}{composition theorem}

%%% Local Variables:
%%% mode: latex
%%% coding: latin-1
%%% TeX-master: "quantum-security"
%%% End:
%%% Local IspellDict: british

\end{appendix}

\newpage
\def\@BIBLABEL#1{[#1]}
\let\oldthebibliography\thebibliography
\def\thebibliography#1{\oldthebibliography{#1\ }}
\oldindex{references}
\bibliographystyle{alpha}
\bibliography{quantum-security}

\newpage
\oldindex{index}
\label{index}
\input{quantum-security.ind}

\end{document}